\documentclass[twocolumn]{aastex631}

\usepackage{appendix}
\usepackage{natbib}
\usepackage{tabularx}
\usepackage[normalem]{ulem}
\usepackage{graphicx}
\usepackage{topcapt}
\usepackage{booktabs}
\usepackage{url}
\usepackage{longtable}
\usepackage{array}
\usepackage{subfigure}
\usepackage{multirow}
\usepackage{color,subfigure}
\usepackage[T1]{fontenc}
\usepackage{academicons}
\usepackage{xcolor}
\usepackage{hyperref}

\usepackage{ifthen}

\usepackage{amsmath}           

\newcommand{\isarxiv}{1}

\newcommand{\RNum}[1]{\uppercase\expandafter{\romannumeral #1\relax}}

\newcommand{\hbeta}{H{$\beta$}}

\def\FeII{Fe\,{\sc ii}}

\def \OIII {[O\,{\sc iii}]}

   \font\sevenrm=cmr7 scaled 1000

\newcommand*{\rom}[1]
{\expandafter\@slowromancap\romannumeral #1@}

\newcommand{\snu}{\affil{Department of Physics \& Astronomy, Seoul National University, Seoul 08826, Republic of Korea: jhwoo@snu.ac.kr}}

\begin{document}

\title{Revisiting the \hbeta\ Size--Luminosity Relation Using a Uniform Reverberation-Mapping Analysis}

\author[0000-0002-2052-6400]{Shu Wang} \snu

\author[0000-0002-8055-5465]{Jong-Hak Woo} \snu

\begin{abstract}

We revisit the relation between active galactic nucleus (AGN) broad-line region (BLR) size and luminosity 
by conducting a uniform \hbeta\ reverberation-mapping analysis for 212 AGNs with archival light curves. Our analysis incorporates three different lag measurement methods, including the interpolated cross-correlation function (ICCF), {\tt JAVELIN}, and {\tt PyROA}, alongside a consistently defined lag searching window and an alias removal procedure. We find that ICCF, albeit with larger uncertainties compared to other methods,  is the most reliable method based on our visual inspection of the matches between \hbeta\ and the shifted continuum light curves. Combining this sample with the 32 AGNs from the Seoul National University AGN Monitoring Project, we obtain the best-fit relation between the BLR size ($R_{\rm BLR}$) and the continuum luminosity at 5100\AA\ ($L_{5100}$) with a slope significantly flatter than 0.5. By selecting a subsample of
157 AGNs with the best-quality lag measurements using a set of quantitative criteria and visual inspection, we find a consistent slope and a slightly decreased intrinsic scatter. We further investigate the effect of luminosity tracers, including  $L_{5100}$, \hbeta\ luminosity ($L_{\rm H\beta}$), \OIII\ luminosity ($L_{\rm [O\,\RNum{3}]}$), and 2 to 10 keV hard X-ray luminosity ($L_{\rm 2\text{--}10\,keV}$).  We find that sub-Eddington and super-Eddington AGNs exhibit systematic offsets in the $R_{\rm BLR}$--$L_{5100}$ and $R_{\rm BLR}$--$L_{\rm H\beta}$ relation, but are comparable in the $R_{\rm BLR}$--$L_{\rm [O\,\RNum{3}]}$ and $R_{\rm BLR}$--$L_{\rm 2\text{--}10\,keV}$ relation. We discuss the potential causes for these different deviations when employing different luminosity tracers.
 
\end{abstract}


\section{Introduction}

The scaling relation between the mass of supermassive black holes (BHs) and the properties of their host galaxies (e.g., the bulge mass and stellar velocity dispersion) is 
one of the key ingredients in our understanding of galaxy evolution \citep{Magorrian98, Ferrarese00, Gebhardt00, Marconi03, Haring04, Gultekin09, Woo10}.
Exploring the evolution of this relation
requires accurate mass determination of distant BHs. However, the traditional method based on stellar/gas dynamics in the BH vicinity becomes impractical due to the insufficient spatial resolution of current observing facilities. 
One alternative approach is reverberation mapping  \citep[RM; e.g.,][]{Blandford82,Peterson93}, which measures the time delay between the continuum variability and the response of broad emission lines, thereby gauging the size of the broad-line region (BLR; $R_{\rm BLR}$) in active galactic nuclei (AGNs). By combining $R_{\rm BLR}$ and the velocity of BLR gas measured from the line widths, the BH mass can be determined under the virial assumption, i.e., BLRs are virialized and dominated by the gravitational force of BHs.

Early studies of RM unveiled a scaling relationship between $R_{\rm BLR}$ and continuum luminosity at 5100\AA\ ($L_{5100}$) 
\citep[e.g.,][]{Kaspi00,Bentz09b}.
Using a sample of $\sim50$ local AGNs with $L_{5100}$ well accounted for host contamination, 
\citet{Bentz13} presented a tight $R_{\rm BLR}$--$L_{5100}$ relation with a small intrinsic scatter of 0.19 dex, and a slope of 0.533$^{+0.035}_{-0.033}$.
Their slope was consistent with the expectation from a simple photoionization model. Such a relation significantly simplified the estimation of BH mass, by requiring only single-epoch luminosities and line widths \citep[e.g.,][]{Woo02, Vestergaard06, Shen12, DallaBonta20}. While a variety of emission lines were utilized for AGNs at different redshifts, the \hbeta\ $R_{\rm BLR}$--$L_{5100}$ relation has been best studied with the local RM sample, providing the foundation of all other virial BH mass estimators beyond the local universe.

However, more recent RM studies, which extended to a broader range of both luminosities 
\citep[e.g.,][]{Grier17b,Li21,Woo24} and Eddington ratios \citep[e.g.,][]{Wang14,Du14,Du15,Du16,Du18,Hu21}, reported a significant increase in the scatter of the $R_{\rm BLR}$--$L_{5100}$ relation.  
In particular, super-Eddington AGNs 
show a significant offset from the $R_{\rm BLR}$--$L_{5100}$ relation of sub-Eddington AGNs.  
By adding 32 AGNs with moderate-to-high luminosity, 
\citet{Woo24} found that the current $R_{\rm BLR}$--$L_{5100}$ relation is characterized by a shallower slope of $\sim0.4$ and a larger intrinsic scatter of 0.23 dex  than
that of \citet{Bentz13}. In addition, the independent observation from GRAVITY confirmed the shallower slope of $R_{\rm BLR}$--$L_{5100}$ relation based on a small sample of AGNs with spectro-interferometric observations \citep{Gravity24}.

For understanding the systematic trends in the relation, \citet{Du_Wang19} reported a strong correlation between the deviation from $R_{\rm BLR}$--$L_{5100}$ relation and the \FeII\ to \hbeta\ flux ratios, which is commonly used as an indicator of accretion rate \citep[e.g.,][]{Shen14}. It means that the higher the accretion rate is, the more the BLR size is shortened, which could be explained by the self-shadowing effect of the slim accretion disk \citep{Wang14a}. The inner part of the slim disk is puffed-up and partially obscures the ionizing photons that illuminate the BLRs, resulting in the overall shortening of $R_{\rm BLR}$. \citet{Woo24} demonstrated that the flattening of the relation is related to the higher average Eddington ratios at higher-luminosity bins. In addition, \citet{Fonseca-Alvarez20} suggested that the deviation likely depends on the UV/optical spectral energy distribution, i.e., the ratio between the optical and the ionizing luminosity.  By using the ratio between \OIII\ and \hbeta\ as the proxy for UV ionizing photons, they demonstrated that the changing in UV to optical SEDs may also contribute in the offset and scatter of the $R_{\rm BLR}$--$L_{5100}$ relation.

While our understanding of the $R_{\rm BLR}$--$L_{5100}$ relation has significantly deepened with the growing RM sample, it is important to note that  the \hbeta\ lags reported in the literature were not measured uniformly, which could introduce extra scatter that complicates the comprehension. Some lags were measured based on the interpolated cross-correlation function (ICCF)  \citep{Peterson98}, while others were based on {\tt JAVELIN} \citep{Zu11}. These two approaches can have significant difference in the lag uncertainty estimation \citep{Li-J19,Yu20}, and their lags can also be inconsistent \citep[e.g.,][]{Woo24}. Thus, it is of importance to conduct a uniform lag analysis for testing the effect of different lag measurement methods.  In addition, there is a range of the reliability of the reported lags in the literature, owing to the diverse quality of light curve data and adopted analysis.
It is important to test whether the observed scatter of $R_{\rm BLR}$--$L_{5100}$ relation can change with a more rigorously selected sample with the best quality lag measurements.

In this paper, we perform a uniform analysis to re-measure \hbeta\ lag for a large sample of the previously reverberation-mapped AGNs in the literature, and revisit the $R_{\rm BLR}$--luminosity ($L$) relation using different luminosity tracers. We summarize the AGN sample and light curve data collection in \S \ref{sec:sampleanddata}.  In \S \ref{sec:lag}, we investigate the influence of different lag measurement methods and lag qualities on $R_{\rm BLR}$--$L_{5100}$ relation.  In \S \ref{sec:luminositytracers}, we study impact of the different luminosity tracers and discuss our current understanding of the $R_{\rm BLR}$ -- $L$ relation in \S5.   In \S \ref{sec:summary}, we give a brief summary of our findings.
Throughout this paper, we use the $\Lambda$CDM cosmology, with $H_0=72.0$, and $\Omega_{m}=0.3$.

\section{Sample and data collection} \label{sec:sampleanddata}

\subsection{Sample}

The sample used in work consists of two parts. First, we collected all AGNs with \hbeta\ RM measurements published before 2022 (literature master sample), where we found $\sim280$ individual lag measurements. Second, we incorporated the lags of 32 moderate-to-high-luminosity AGNs from the Seoul National University AGN Monitoring Project \citep[SAMP;][]{Woo19,Woo24}. 

For the literature master sample, we collected the \hbeta\  and continuum light curves for each object from the original papers to perform a uniform lag analysis. In this process,  we removed objects for which the two light curves were not available. In some cases, we applied small adjustments to the archival light curves whenever required, e.g., excluding outliers and inter-calibrating between different bands or telescopes, following the guidance from the original papers. We also collected the lags reported in the literature ($\tau_{\rm literature}$) for comparison purposes.

To investigate the $R_{\rm BLR}$--$L_{5100}$ relation,  we also collected the host-corrected AGN continuum luminosity $L_{5100}$. The host contamination can be estimated based on space/ground-based image decomposition \citep[e.g.,][]{Bentz13}, spectral decomposition or empirical relations \citep[e.g.,][]{Shen11}. We excluded 38 objects, for which any host galaxy correction is not available. We mainly collected the $L_{5100}$ from the database of \citet{Bentz15} and supplemented this database with the additional new measurements from \citet{DallaBonta20} as well as other individual papers. We revised these luminosities with the same cosmological parameters, except for very nearby AGNs, for which the direct distance measurements were adopted.  

In addition we excluded 28 unreliable lag measurements due to the under-sampling or low variability as reported by the previous works \citep{Bentz15}. Thus, we finalized the remaining 212 AGNs with \hbeta\ RM measurements as the literature parent sample for the following lag analysis. For clarity, we assigned an ID to each AGN in this sample (see Table \ref{tab:Lag_measurement} in the Appendix for details).

\begin{table}[htbp]
    \centering
    \caption{Sample Collection}
    \begin{tabular}{p{0.38\textwidth} c}
    \hline \hline
    Sample      &  $N_{\rm objects}$ \\ \hline 
    Literature master sample & 280 \\
    ~~~~Light curves not available &  2 \\
    ~~~~Host-subtracted luminosity not provided & 38 \\
    ~~~~Previously identified as un-reliable lags & 28 \\
    Literature parent sample & 212 \\
        \hline
    SAMP  & 32 \\
    \hline
    Total sample  &  244 \\
     \hline
    \end{tabular}
    
    \label{tab:sample_stat}
\end{table}

Then we combined the literature parent sample with the 32 moderate-to-high-luminosity AGNs from SAMP. The measured lags based on ICCF and {\tt JAVELIN} are presented  by \citet{Woo24}, along with the host-corrected $L_{5100}$ based on spectral decomposition. Note that we adopted a similar methodology and lag quality assessment as \citet{Woo24}. The final total sample contains 244 AGNs (see Table \ref{tab:sample_stat}).

\subsection{Other Properties}

We collected several spectral properties for the literature parent sample, including broad \hbeta\ luminosity ($L_{\rm H\beta}$), \OIII\ luminosity ($L_{\rm [O\,\RNum{3}]}$), the line dispersion ($\sigma_{\rm line}$) measured from the root-mean-square (rms) spectra, and the full-width-at-half-maximum (FWHM) measured from the mean spectra.
The $L_{\rm H\beta}$ is a direct product of BLR ionization, thus is a good indicator of the ionizing luminosity. It is also less subject to the difficulty in removing the host galaxy contribution \citep{DallaBonta20}.  The $L_{\rm [O\,\RNum{3}]}$ may reflect the amount of higher energy photons because \OIII\ has an ionization potential of $\sim$ 55 eV.  
For the majority of our sample, these measurements were collected from \citet{Du_Wang19}; the properties for the Sloan Digital Sky Survey -- RM (SDSS--RM) were collected from \citet{Shen19}; the properties of SAMP AGNs were measured by this work.  

We also collected the velocity dispersion measured from rms spectra ($\sigma_{\rm line, rms}$) for calculating the BH mass and the Eddington ratio ($\lambda_{\rm Edd}$).  The $\sigma_{\rm line, rms}$  is primarily used for $M_{\rm BH}$ calculation,  because it is generally more consistent with the virial assumption \citep{Peterson04, Park12, Wang20} and less subject to orientation effect \citep[e.g.,][]{Collin06,Wang19,Yu20b}.  For objects with no available  $\sigma_{\rm line, rms}$, the FWHM$_{\rm mean}$ were adopted. We adopted a single representative virial coefficient $f=4.47$ for  $\sigma_{\rm line, rms}$ and $f=1.12$ for FWHM \citep{Woo15}.

\section{Investigation of uniform lag measurement} \label{sec:lag}

\subsection{Lag measurement}

\begin{figure*}

\ifthenelse{\isarxiv=1}{
\includegraphics[width=0.99\textwidth]{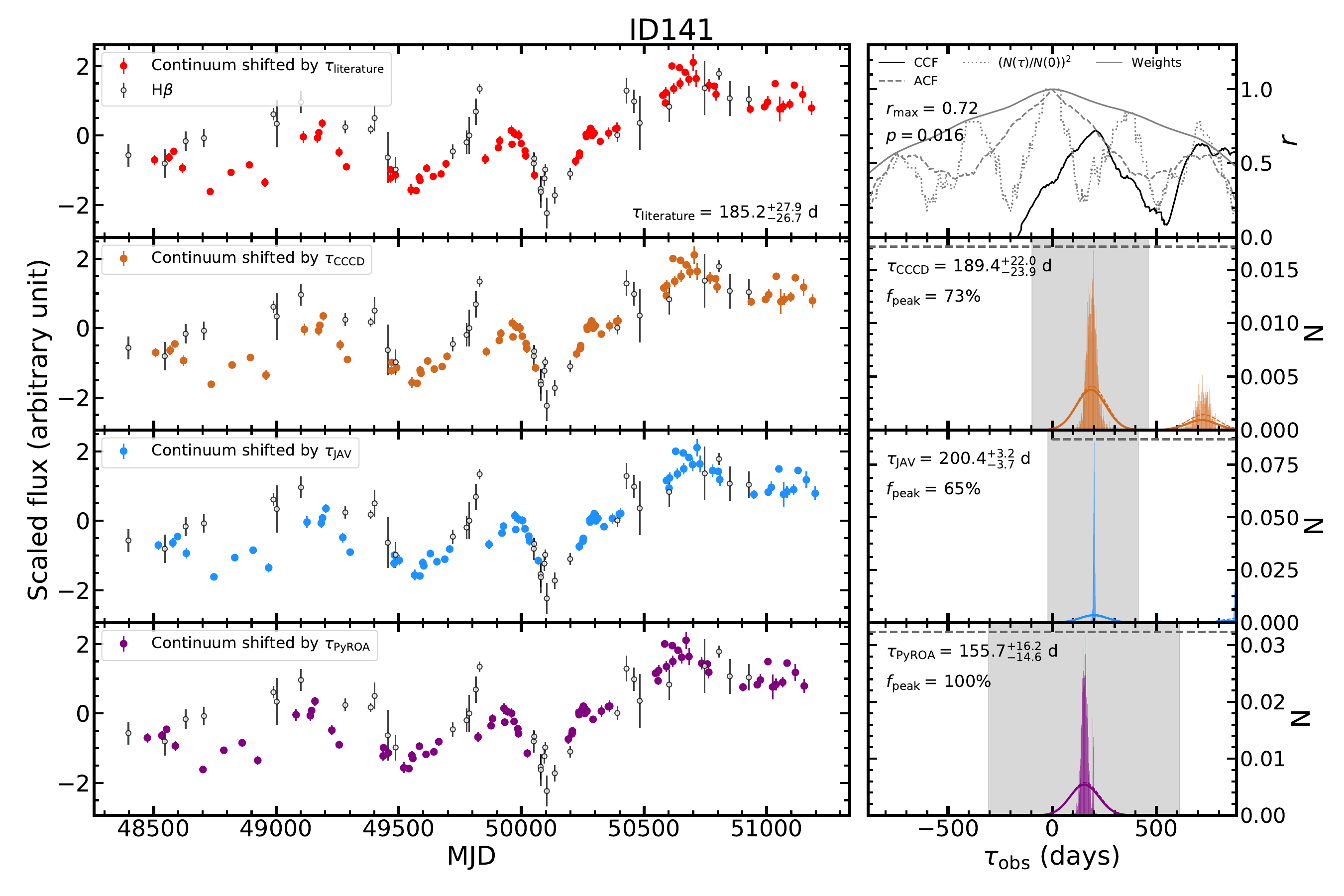}
\caption{Example of lag measurements based on ICCF, {\tt JAVELIN}, and {\tt PyROA} for ID141.  Left: the \hbeta\ light curve (black open circles) is compared to the continuum light curve shifted by $\tau_{\rm literature}$ (top panel; red circles), $\tau_{\rm ICCF}$ (second panel; brown circles), $\tau_{\rm JAV}$ (third panel; blue circles), and $\tau_{\rm PyROA}$ (bottom panel; purple circles), respectively.  Right:  Top panel: the CCF (black solid line) and weighting functions, including the ACF (grey dashed line), overlapping function (grey dotted line), as well as the final weights (grey solid line; see \S \ref{sec:alias_removal} for details). The $r_{\rm max}$ and $p$-value are also labeled at the left, which describe the correlation strength and the significance of the correlation, respectively (see \S \ref{sec:lag_quality} for more details). Second to bottom panels: the posterior distribution of $\tau_{\rm ICCF}$ (second panel),  $\tau_{\rm JAV}$ (third panel), and $\tau_{\rm PyROA}$ (bottom panel).  The dashed and solid curves represent the smoothed unweighted and weighted distribution, respectively. The grey shaded area illustrates  the range of the primary peak. The lag measurement as well as the fraction ($f_{\rm peak}$) of the posterior distribution within the primary peak is labeled at the left. The horizontal dashed lines at the top represent the lag searching window adopted for this object. Complete figures for each object are presented online in their entirety. }
\label{fig:lag_aliasremoval_example}
}{
\figsetstart
\figsetnum{1}
\figsettitle{ICCF, JAVELIN, and PyROA lag measurements for 212 objects in the literature parent sample.}

\figsetgrpstart
\figsetgrpnum{1.141}
\figsetgrptitle{ID141}
\figsetplot{141}
\figsetgrpnote{ICCF, JAVELIN, and PyROA lag measurements.}
\figsetgrpend

\figsetend

\figurenum{1}
\plotone{ID141.pdf}
 \caption{Example of lag measurements based on ICCF, {\tt JAVELIN}, and {\tt PyROA} for ID141.  Left: the \hbeta\ light curve (black open circles) is compared to the continuum light curve shifted by $\tau_{\rm literature}$ (top panel; red circles), $\tau_{\rm ICCF}$ (second panel; brown circles), $\tau_{\rm JAV}$ (third panel; blue circles), and $\tau_{\rm PyROA}$ (bottom panel; purple circles), respectively.  Right:  Top panel: the CCF (black solid line) and weighting functions, including the ACF (grey dashed line), overlapping function (grey dotted line), as well as the final weights (grey solid line; see \S \ref{sec:alias_removal} for details). The $r_{\rm max}$ and $p$-value are also labeled at the left, which describe the correlation strength and the significance of the correlation, respectively (see \S \ref{sec:lag_quality} for more details). Second to bottom panels: the posterior distribution of $\tau_{\rm ICCF}$ (second panel),  $\tau_{\rm JAV}$ (third panel), and $\tau_{\rm PyROA}$ (bottom panel).  The dashed and solid curves represent the smoothed unweighted and weighted distribution, respectively. The grey shaded area illustrates  the range of the primary peak. The lag measurement as well as the fraction ($f_{\rm peak}$) of the posterior distribution within the primary peak is labeled at the left. The horizontal dashed lines at the top represent the lag searching window adopted for this object. Complete figures for each object are presented online in their entirety. }
\label{fig:lag_aliasremoval_example}
}

\end{figure*}

Two lag measurement methods are primarily adopted in the literatures, i.e., ICCF \citep[e.g.,][]{Gaskell_Peterson87,Peterson98} and  {\tt JAVELIN} \citep{Zu11}.  ICCF is a non-parametric, model independent method,  while {\tt JAVELIN} is  based on the damped random walk model \citep{Kelly09}. Detailed comparison between these two methods have been performed based on simulated light curves \citep{Li-J19,Yu20}.  The results show that both methods are generally robust. However, ICCF tends to overestimate the lag uncertainty than the value determined from random trails, while {\tt JAVELIN} is more susceptible to the underestimation of the light curve flux uncertainties. In this work, we compare these methods using archival light curves from real observation. In addition, we supplement our analysis with a third approach {\tt PyROA} \citep{Donnan21},  which is recently developed based on a different methodology and is also effective in recovering lags \citep{Shen23,Khatu23}. In the following, we briefly outline the methodology of these methods and the parameter setting. 

ICCF measures the cross-correlation function (CCF) using linear interpolation. Specifically, at each step $\tau$ of a lag searching array, one light curve is shifted by $\tau$ and linearly interpolated on the time grid of the other. The Pearson cross-correlation coefficient $r$ is calculated between the interpolated first light curve and the second light curve. This process is repeated over the entire lag searching array, providing $r$ as a function $\tau$, which is called as ICCF. Exchanging the order of the first and second light curve results in two ICCFs, and the final ICCF is the average of these two.

The ICCF lag measurement can be derived from either the peak or the centroid  of ICCF, with the latter considered as the more robust measure \citep{Peterson04} which is adopted in this work.  Specifically, the ICCF centroid is calculated using points with $r\geq0.8r_{\rm max}$ \citep{Peterson04} around the $r_{\rm max}$.  By employing the Flux Randomization (FR) and Random Subset Selection (RSS) methods, we generate 5000 realizations of the light curves and obtain the sampling distribution.  The details of the FR and RSS approach are described in \citet{Peterson98} and implemented by the package {\tt PyCCF} \citep{Sun18}. In the following text, we ignore the difference between posterior distribution and sampling distribution, while acknowledging that the ICCF outputs are actually different from posterior distribution based on Markov Chain Monte Carlo  (MCMC) techniques.

{\tt JAVELIN} utilizes the damped random walk (DRW) model \citep{Kelly09} to simulate and interpolate light curves. It describes the emission-line light curve as a  shifted, smoothed version of the continuum light curve \citep{Zu11} using a top-hat transfer function. The best-fit parameters of the DRW model and the transfer function are determined using MCMC techniques. In this work, we utilize the {\tt JAVELIN} v0.33 and adopt the $n_{\rm burn}=n_{\rm walk}=100$ and $n_{\rm chain}=300$ as the MCMC parameters.  
 
{\tt PyROA} utilizes the running optimal average (ROA) to model light curves  \citep{Donnan21} which represents an inverse-variance weighted average of all data points.  It is an empirical method without assumption of the DRW model.  
The fitting is performed on the stacked light curves which are merged from the continuum and the shifted and scaled \hbeta\ light curves. The ROA model is multiplied with a Gaussian window function with a width of $\Delta$ to suppress influence from distance points.  The best-fit model is derived by balancing the goodness of the fit (i.e., $\chi^2$) and the effective number of parameters which is a function of $\Delta$. The model parameters are determined using Bayesian analysis.  In this work, we adopted $N_{\rm sample}=10000$ in the MCMC calculation and [5.0, 30.0] days as the $\Delta$ range.  In a small number of cases, the $\Delta$ range is adjusted to prohibit overfitting with a small $\Delta$ value. As for the initial guess of the lag for {\tt PyROA}, we utilize the ICCF lag as an input.

Next, we describe how we define the lag searching windows. 
In principle, a longer lag searching window is able to search for more possibilities, while in practice it results in more aliases and even a failure in recovering the real lag. 
In this work, the light curve from campaigns with different length focusing on objects with different lags, which makes it difficult to find a uniform optimal solution for each object. After testing several different options based on the time baseline length, we decide to use the lag searching window as $\pm0.33$ times of the  baseline for ICCF and {\tt PyROA}, which is similar to what is adopted by \citet{Woo24}.  For {\tt JAVELIN}, we use [0, $+0.33$] times of the baseline as the searching window because we find that  symmetric windows often cause negative lag measurements that are not physically meaningful.   
Most of these adopted searching windows are larger than the expected lags from the  $R_{\rm BLR}$--$L_{5100}$ relation of \citet{Bentz13}. However, there are 15 objects whose expected lags from \citet{Bentz13} are larger than our lag searching windows. For these targets, we additionally use $\pm$0.5 times of  the baseline as the searching window and found that the ICCF result doesn't change.  We argue that employing larger windows can result in lags that are less reliable, because such lags lead to smaller overlap between the shifted continuum and \hbeta\ light curves using the lag.

\subsection{Alias removal} \label{sec:alias_removal}

Next, we describe our methodology of removing aliases in the posterior distributions. All these three approaches can show multiple peaks in their posterior distribution due to the limited time sampling and baseline length, quasi-periodic variability, and etc. \citep{Grier17b, Homayouni20}. The situation can be more severe when there are large seasonal gaps in the light curves. In these cases, applying a weighting function to suppress and remove the alias is necessary. 

We first smooth the posterior distribution using a Gaussian kernel whose width is empirically determined to be 0.075 times of the baseline length of the continuum light curves. This value is determined by visual inspection of the smoothed posterior distribution. Then the smoothed posterior distribution is multiplied with the weights as demonstrated in Figure \ref{fig:lag_aliasremoval_example}.  Similar to those applied by \citet{Grier19}, the weights consists of two components: the auto-correlation function (ACF) the overlapping function. The ACF reflects the rapidity of the variability, where a narrow ACF reflect rapid variability where the lag is more easily affected by the seasonal gap. The overlapping function is calculated by $P(\tau)=(N(\tau)/N(0))^2$, where $N(\tau)$ is the number of overlapping points between the two light curves after shifting the continuum light curve with $\tau$. 
At specific $\tau$, the overlapping points are defined as those \hbeta--shifted continuum epoch pairs for which the time difference is less than twice the median cadence of the continuum light curves. We follow the approach by \citet{Grier17b} to clip ACF  at its first zero-point (if any) surrounding $\tau=0$, and the final weights is the convolution between the clipped ACF and the overlapping function $P(\tau)$.

The primary peak is defined as the highest peak in the weighted smoothed posterior distribution, and its range is determined as the two adjacent local minima around the primary peak or the searching window if  no local minimum is found. The final lags are determined from the unweighted posterior distribution that are inside the primary peak range.

\renewcommand\thefigure{\arabic{figure}} 
\setcounter{figure}{1}
\begin{figure*}[htbp]
    \centering
    \includegraphics[width=0.99\textwidth]{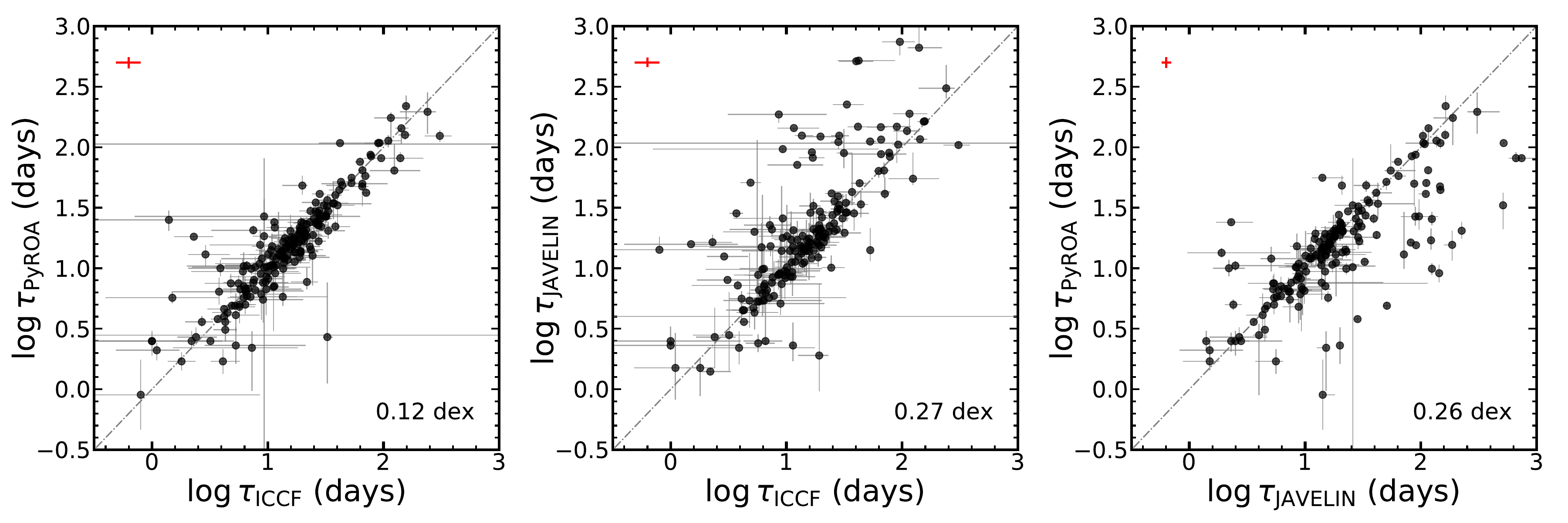}
    \caption{Comparison between $\tau_{\rm ICCF}$,  $\tau_{\rm JAV}$, and $\tau_{\rm PyROA}$. The grey dotted dashed line represents the 1:1 relation.  The  red crosses at the top left corner indicate the median uncertainties of x- and y-axis. The scatter  in 16-to-84th percentile range is shown at the bottom right corner. }
    \label{fig:ICCF_JAVELIN}
\end{figure*}

\subsection{Detrending}

We explore the effect of removing the long-term trend, a process known as detrending,  on the lag measurements. The presence of a long-term trend can skew the lag measurement \citep{Peterson95}, and moreover, the different long-term trends in the continuum and \hbeta\ light curves can reduce the correlation strength \citep{Zhang19,Li20}. We perform a test to detrend both the continuum and \hbeta\ light curves using linear function, where the best-fit linear trend is subtracted in both light curves. We perform ICCF calculation on the detrended light curves. While for most objects the correlation strength  becomes weaker, $\sim$20 objects show stronger correlation strength. We find that most of these objects show consistent lag measurements before and after detrending. Only five of them show inconsistent lags at 3$\sigma$ level, e.g.,  3C 273 \citep{Zhang19}. Based on our visual inspection, we adopt the lags with the detrending when the $r_{\rm max}$ is improved by 0.1 and the lags are inconsistent at 3$\sigma$ level before and after the detrending.

\subsection{Lag comparison} \label{sec:lag_comparison}

We present the lag measurements of the literature parent sample based on the three methods in Table \ref{tab:Lag_measurement}.  In Figure \ref{fig:ICCF_JAVELIN}, we compare the the lags from these methods (i.e., $\tau_{\rm ICCF}$,  and $\tau_{\rm JAVELIN}$, and $\tau_{\rm PyROA}$).  In the comparison, we remove those lags that are not converged or affected by the boundary of lag searching windows. We find that (1) $\tau_{\rm ICCF}$ and $\tau_{\rm PyROA}$ are mostly consistent with each other; (2) while for most objects $\tau_{\rm JAVELIN}$ is consistent with other two lags, there are many outliers. 

To examine the outliers and determine which measurements might be problematic, we shift the continuum light curves according to the lags of each of these three methods,  and align the flux based on the median and standard deviation (Figure \ref{fig:lag_aliasremoval_example}).  Visual inspection of these figures suggests that the discrepancy is primarily associated with inaccurate $\tau_{\rm JAVELIN}$, while $\tau_{\rm ICCF}$ (and $\tau_{\rm PyROA}$) generally yield a better match between the light curves.  For some objects,  for example, ID057 and ID062, $\tau_{\rm JAVELIN}$ matches the continuum and \hbeta\ light curves in the seasonal gap, leading to almost no overlap between these two.

Similar results have been observed in several monitoring campaigns with multi-year baseline and moderate cadence \citep{Shen23,Woo24}. These works find that the failed $\tau_{\rm JAVELIN}$ often match the two light curves in the seasonal gaps.   This tendency of {\tt JAVELIN} to favor lags close to the length of seasonal gaps may arise from its inadequate handling of underestimated flux uncertainties. When the flux is imprecise and its uncertainty is underestimated,  {\tt JAVELIN}'s best solution tends to minimize the overlap to reduce the mismatch.  
In addition, we find that some $\tau_{\rm JAVELIN}$ are not converged, which is not seen in the case of ICCF and {\tt PyROA}. 
These results suggest that ICCF and {\tt PyROA} are generally more reliable than {\tt JAVELIN}.

\begin{figure}[htbp]
   \centering
   \includegraphics[width=0.45\textwidth]{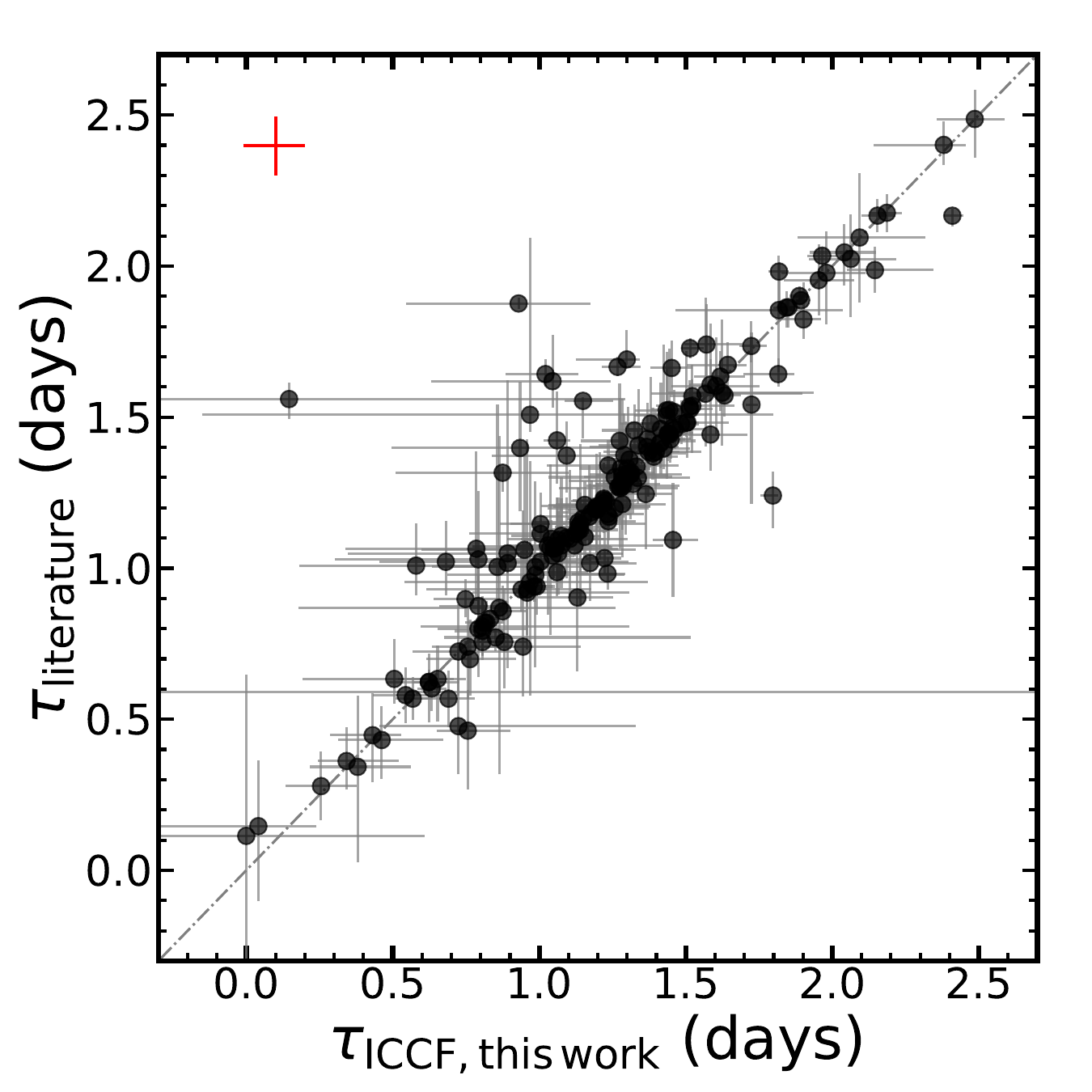} 
   \includegraphics[width=0.45\textwidth]{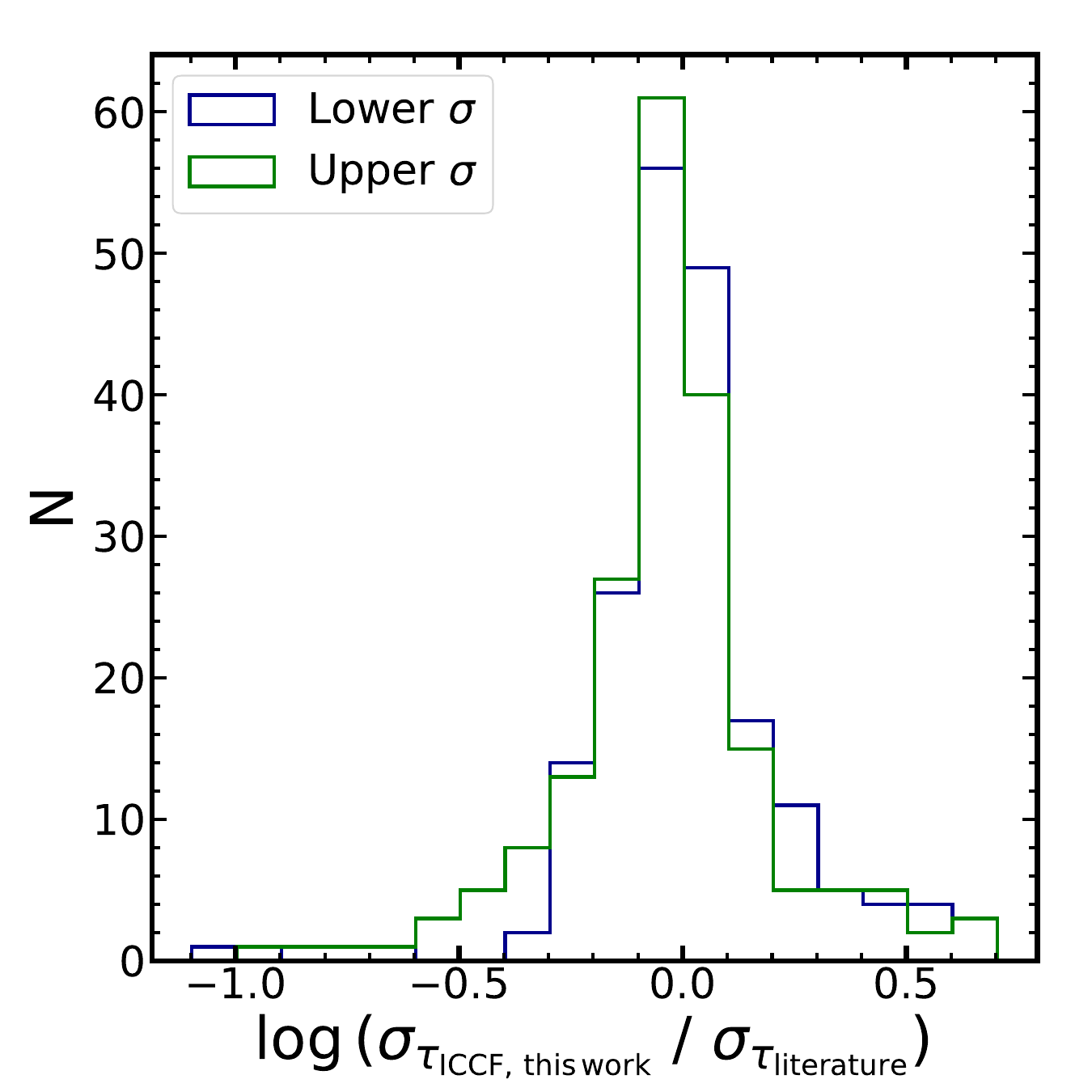}
   \caption{Upper: comparison between the lags originally reported in the literature ($\tau_{\rm literature}$) and the lags measured in this work based on ICCF ($\tau_{\rm ICCF, this\,work}$).  The grey dotted-dashed line represents the 1:1 correspondence.  The red cross at the top-left corner indicates the  log scale average uncertainty of both axes. Lower: distribution of the difference between the lag uncertainty of $\tau_{\rm literature}$ and the uncertainty of $\tau_{\rm ICCF, this\,work}$. Lower and upper uncertainties are shown using blue and green color, respectively. }
   \label{fig:ICCF_Paper}
\end{figure}

Despite that $\tau_{\rm ICCF}$ are generally reliable,  there are some cases where  $\tau_{\rm JAVELIN}$  or $\tau_{\rm PyROA}$ are more accurate, presenting better matches between \hbeta\ and the shifted continuum light curves than $\tau_{\rm ICCF}$. Such cases include ID 074, ID 133, and ID 145. A common feature of these objects is that there is a primary non-linear long-term trend, e.g., a bowl-like structure, in their light curves. In this case, their $\tau_{\rm ICCF}$ are sometimes overestimated. On the contrary, the other two lags can successfully match the feature in the light curves. It implies the need for more complex detrending to obtain unbiased ICCF lag measurements for such objects.

For lag uncertainties, we find that the uncertainty of $\tau_{\rm ICCF}$ is about 2.2 and 2.1 times larger than that of $\tau_{\rm PyROA}$and  $\tau_{\rm JAVELIN}$, respectively. This finding is consistent with the results based on simulated light curves \citep{Li-J19, Yu20}. \citet{Yu20} reports that ICCF typically overestimates the 'true' uncertainty--calculated from random realization of the original light curve--by a factor of approximately 3, while {\tt JAVELIN} only slightly overestimates the 'true' uncertainty by 1.1 times.  As for the comparison between the  $\tau_{\rm PyROA}$  and $\tau_{\rm JAVELIN}$,  the uncertainty of $\tau_{\rm JAVELIN}$ is about 1.1 times larger than that of {\tt PyROA}. 

Combining these results, it seems that $\tau_{\rm PyROA}$ is the best option, given the reliable lag measurements and small lag uncertainties. However, it's important to note that we utilize $\tau_{\rm ICCF}$ as initial values in $\tau_{\rm PyROA}$ calculations. Consequently, it's natural for $\tau_{\rm PyROA}$ to demonstrate better consistency with $\tau_{\rm ICCF}$ than $\tau_{\rm JAVELIN}$.  To assess this effect, we conduct an experiment using $\tau_{\rm JAVELIN}$ as the initial value in {\tt PyROA} calculation. This test reveals that for some objects, the recalculated $\tau_{\rm PyROA}$  shift to more closely align with $\tau_{\rm JAVELIN}$, especially when $\tau_{\rm JAVELIN}$ and $\tau_{\rm ICCF}$ exhibit significant discrepancies. It suggests the $\tau_{\rm PyROA}$ of the current version is dependent on the initial values. Furthermore, for some objects with relatively low cadence, we have to manually adjust the $\Delta$ range in {\tt PyROA} calculations to avoid over-fitting caused by small $\Delta$ values.  Therefore,  $\tau_{\rm ICCF}$ provides the most robust lag measurements.

In the subsequent analysis,  we adopt $\tau_{\rm ICCF}$ as our final lag measurement, despite it having larger uncertainties than the other two methods. Note that  $\tau_{\rm PyROA}$ is also utilized to evaluate how the lag uncertainties may affect the intrinsic scatter of the $R_{\rm BLR}$--$L_{5100}$ relation.  

As a consistency check, we compare our new lag measurements ($\tau_{\rm ICCF}$) with those originally reported in the literature in Figure \ref{fig:ICCF_Paper}. The two measurements are mostly consistent with each other and their uncertainties also show no large difference. This general consistency is expected in the sense that ICCF is the primary method used in the literature. 
There are some objects whose new lag uncertainties are larger than the previously reported values, owing to the larger searching window or a different lag measuring method. While our new measurements are generally consistent with the previously reported ones, we uniformly determine the lag and uncertainty of the large sample, providing a valuable database for future works.

\subsection{Quality Assessment}\label{sec:lag_quality}

\subsubsection{Literature Sample}

In this section, we perform a quality assessment for the literature sample by setting several criteria, which were similarly applied by \citet{Woo24}. We focus on the quality assessment of $\tau_{\rm ICCF}$ as it is adopted as the primary lag measurement. First, we require that the maximum correlation coefficient $r_{\rm max}$ is larger than 0.55. This threshold is determined based on our visual inspection, and is similar to the value used in the literature, e.g., 0.5 or 0.6 \citep[][]{Grier17b,U22,Woo24}. 

Second, we check the significance of the correlation, which is also affected by the variability amplitude and the presence of strong variability features.  To quantify it, we generate random \hbeta\ light curves and study the possibility of these mock light curves having the same or stronger correlation than the observed one. Specifically, we generated 1000 mock \hbeta\ light curves for each object with flux randomized at each point but keep the same sampling pattern as the original  light curves. Then a linear trend that was derived from the fitting of original \hbeta\ light curves was added to these random emission-line light curves. These mock \hbeta\ light curves were then cross-correlated with the observed continuum light curves, and the results were compared with the those from real observation. With 1000 realizations, we calculated the $p$-value
as the fraction of the the mock realization having $r_{\rm max}$ larger than the observed $r_{\rm max}$. The smaller the $p$-value is, the higher the significance of the correlation is. However, there is not a clear threshold for this $p$-value to indicate the significance level. We empirically chose $p$-value equal 0.1 as the threshold.

\begin{figure}
    \centering
    \includegraphics[width=0.47\textwidth]{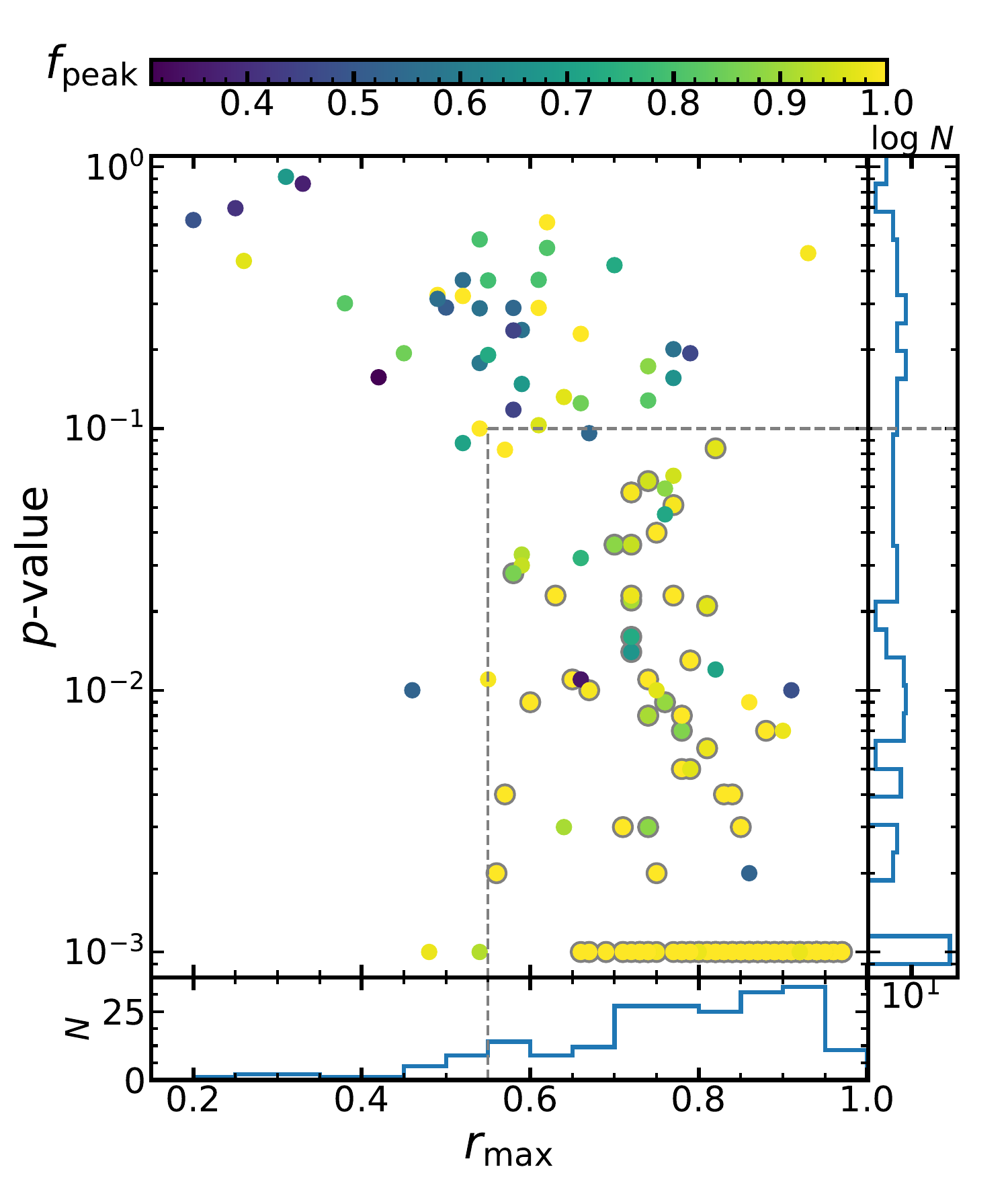}
    \caption{Demonstration of our selection for the best quality lag measurements.  The literature parent sample (212 objects) are plotted on the $r_{\rm max}$--$p$-values plane, and color-coded by the $f_{\rm peak}$ of the $\tau_{\rm ICCF}$.  Objects with $r_{\rm max} >0.55$, $p$-value $\leq0.10$ (as indicated by the grey dashed lines), and $f_{\rm peak}\geq0.6$ are selected as good-quality lag measurements. The histogram of both $r_{\rm max}$ and $p$-values are also displayed in the bottom and right panel, respectively. In addition, we perform visual assessment on the lag quality. Those objects selected based on our visual inspection are denoted with circles with grey edges.  }
    
    \label{fig:selection}
\end{figure}

\begin{table}[]
    \centering
    \caption{Selection of Best-quality Lag Measurements for Literature Sample}

    \begin{tabular}{p{0.38\textwidth} c}
    \hline \hline
    Sample      &  N$_{\rm objects}$ \\ \hline 
    Literature parent sample  &    212 \\
    $\tau_{\rm ICCF} -  \sigma_{\tau_{\rm ICCF}} >0$ & 197  \\
    $r_{\rm max}>0.55$, $p$-value $\leq0.10$, $f_{\rm peak}\geq 0.6$  & 156 \\ 
    Removed by visual inspection   & 22 \\ 
    \hline
    Best-quality literature sample & 134 \\
     \hline
    \end{tabular}
    
    \label{tab:sample_selection}
\end{table}
\begin{figure*}[htbp]
    \centering
    \includegraphics[width=0.89\textwidth]{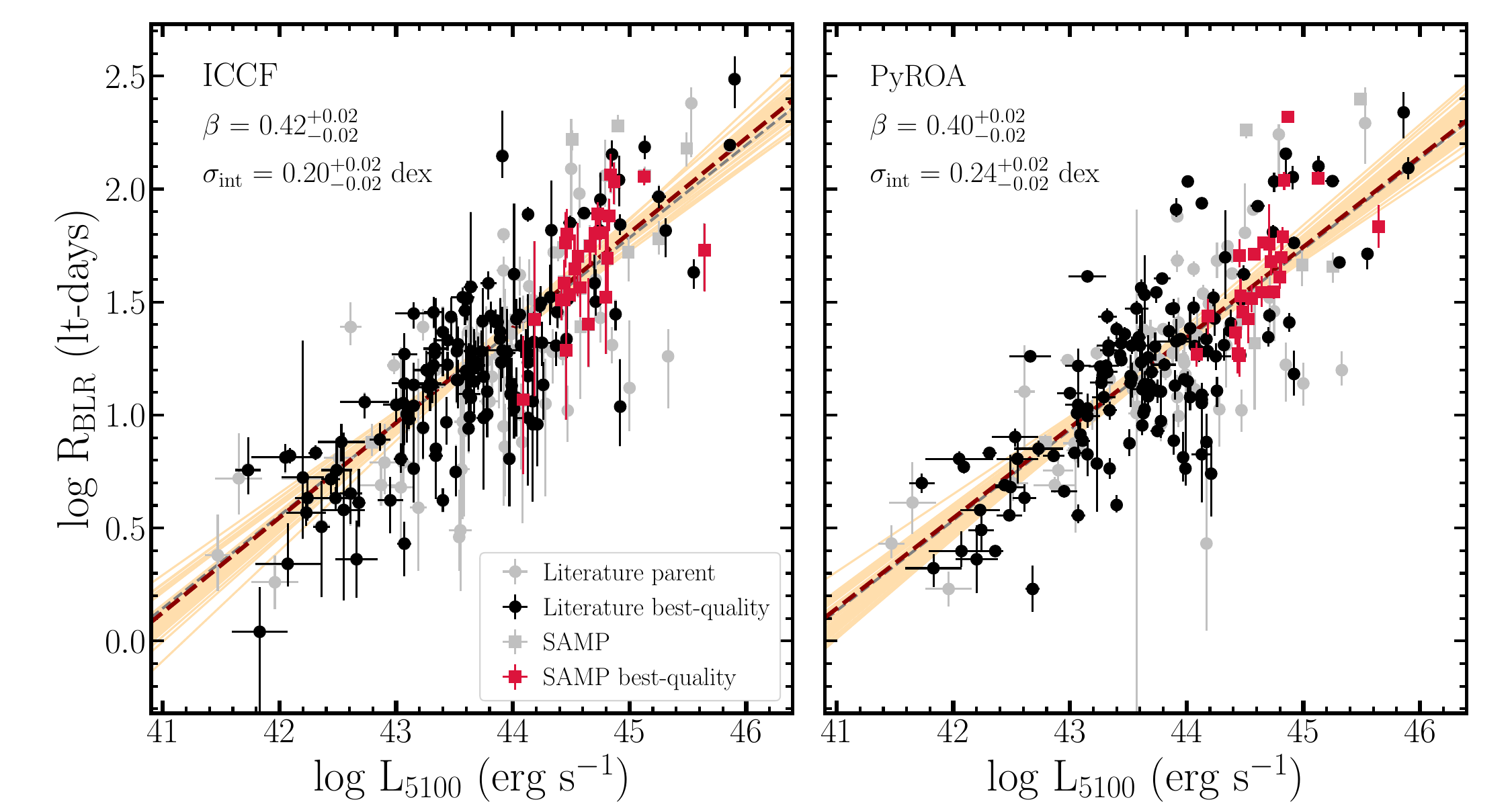}

    \caption{$R_{\rm BLR}$--$L_{5100}$ relation based on ICCF (left) and PyROA (right). The sample presented here includes positive lags  from the  literature parent sample (grey filled circles) as well as the 30 positive lags from SAMP \citep[][grey open circles]{Woo24}. The black filled and open circles represent the best-quality literature sample (see \S \ref{sec:lag_quality} for details) and the good-quality lags selected by \citet[][black open circles]{Woo24}. The grey and brown dashed lines represent the best-fit relations for the combined total and combined best-quality subsample, respectively. The orange lines are 50 randomly drawn realizations from the MCMC samples for illustrating the dispersion of the relation. The best-fit slope $\beta$ and the intrinsic scatter $\sigma_{\rm int}$ are labeled in the legend.   }
    \label{fig:Total_Reliable_ICCF_PyROA}
\end{figure*}

Third, we performed a quality assessment (QA) by visually inspecting the goodness of the match between \hbeta\ and shifted continuum light curves  (based on the $\tau_{\rm ICCF}$), as well as the posterior distribution.  The specific factors we considered include the follows:
\begin{itemize}
    \item   There are obvious variability features (beyond the linear trend) that are clearly matched between \hbeta\ and the shifted continuum light curves.
    \item  The overlapping fraction between \hbeta\ and the shifted continuum light curves is large, at least enough to cover the several variability features, which is necessary to avoid random correlation. The cadence is sufficient to describe the variability features.
    \item  The CCF has a clear peak around the measured $\tau_{\rm ICCF}$.  Note that a flat CCF or a CCF with multiple peak may indicate ambiguous lag measurements.
    \item  The primary peak in $\tau_{\rm cent}$ posterior distribution is well defined, i.e., no other strong peaks outside the primary peak range,  no connected sub-peaks inside the primary peak ranges. The primary peak is not affected by the boundary of the lag searching window.
\end{itemize}

We present the result of our quantitative assessment in Figure \ref{fig:selection} and summarize the number of objects in each selection step in Table \ref{tab:sample_selection}.
Adding additional visual inspection only slightly reduces the sample size. Note that this further selection based on visual inspection does not affect the $R_{\rm BLR}$--$L_{5100}$ relation. 
By combining the quantitative criteria and the visual inspection, we finally obtain the best-quality literature sample, which consists of 134 lag measurements.

\subsubsection{Best-quality Lags From SAMP}

\citet{Woo24} measured 32 \hbeta\ lags from SAMP,  in which 30 objects show significantly positive lags ($1\sigma>0$) based on $\tau_{\rm ICCF}$.  SAMP $\tau_{\rm ICCF}$ are measured using a searching window of $\sim$30\% of the baseline ([-600, 600] days), which typically exceeds twice of the expected lags predicted from the canonical $R_{\rm BLR}$--$L_{5100}$ relation \citep{Bentz13}. They applied the same alias removal procedure as we described above. However, \citet{Woo24} didn't conduct {\tt PyROA} analysis. Consequently,  we measure $\tau_{\rm PyROA}$ for these AGNs, adhering the same parameter setting previously mentioned, such as the adoption of $\tau_{\rm ICCF}$ as initial values. These $\tau_{\rm PyROA}$ measurements  are summarized in the Appendix.  

\citet{Woo24} performed a quality assessment on $\tau_{\rm ICCF}$ and selected a subsample with good-quality lag measurements, using three criteria: (1) $r_{\rm max}>0.6$; (2) $p(r_{\rm max}) \leq 0.20$; (3) $f_{\rm peak}\geq0.6$. Note that their $p(r_{\rm max})$ is different from the $p-$value calculated in this paper. To calculate the $p-$value, \citet{Woo24} utilizes the DRW model to generate the random light curves \citep[][Guo et al., in preparation]{U22}, obtaining the statistical significance of the correlation of two red-noise like signals. In this work we generate the $p$-value based on white-noise-like random light curves, which is a more conservative approach compared to $p(r_{\rm max})$. Detailed comparison between these two approaches is beyond the scope of this work and will be explored in the future.

For consistency,  we apply the same criteria for SAMP AGNs as we do for the literature sample, i.e., relaxing the first threshold to $r_{\rm max}>0.55$ and recalculating the $p$-values based on our methodology. We find that all the good-quality lag measurements from SAMP satisfy these two requirements. The third criterion is the same between this work and \citet{Woo24}.  Upon our visual inspection, we also confirm the high-quality lags identified by \citet{Woo24}.

Note that we disquality two objects from the best-quality measurements. One object is removed by requiring $\tau_{\rm ICCF}$ to be 1$\sigma>0$, and the other object is removed because its $\tau_{\rm JAVELIN}$ is preferred than $\tau_{\rm ICCF}$ \citep{Woo24}.  Thus, we finalize 23 AGNs from SAMP as best-quality measurements for the following analysis.

\subsection{Effect of the Uncertainty and Quality of Lag Measurements on the $R_{\rm BLR}$--$L_{5100}$ Relation}  \label{sec:R-L_relation_ICCF_PyROA}

 In this section, we investigate the  $R_{\rm BLR}$--$L_{5100}$ relation using the total sample and the best-quality subsample, respectively, to investigate the effect of the lag measurement quality.
Note that we only utilize the positive lags that are $1\sigma>0$ in the fitting. We fit the $R_{\rm BLR}$--$L_{5100}$ relation using the following form:
\begin{equation}
    y = \alpha + \beta (x - x_0)
\end{equation}
where $x$ is an independent and $y$ is a dependent variable. In the expression, the independent variable $x$ is normalized according to its average $x_0$, which equals to 43.89 for log$L_{5100}$ in our sample.  To determine the best-fit parameters,  we utilize the MCMC analysis implemented by {\tt emcee}\footnote{\hyperlink{https://github.com/dfm/emcee}{https://github.com/dfm/emcee}} package \citep{Foreman-Mackey13}. where the likelihood $\mathcal{L}$ is expressed by:
\begin{equation}
    {\rm ln} \mathcal{L}^2 = - \frac{1}{2}\sum^{N}_{i=1}{\frac{(y_i-m_i)^2}{s_i^2} + {\rm ln} (2\pi s_i^2)}
\end{equation}
where the $y_i$ and $m_i$ are observation and model predictions of the $i$th object. The $s_i$ is expressed as

\begin{equation}
    s_i^2 = (\beta x_{\rm err})^2 + y_{\rm err}^2 + \sigma_{\rm int}^2 
\end{equation}
where the last term $\sigma_{\rm int}$ is added to represent the intrinsic scatter of the relation. This is similar to other regression methods in the literature, e.g., {\tt LINMIX} \citep{Kelly07}. In {\tt emcee} calculation, we adopt flat prior distributions within reasonable parameter space. The best-fit values are determined as the median of the posterior distribution, with uncertainties represented by the range between the 16th and 84th percentiles.

 In Figure \ref{fig:Total_Reliable_ICCF_PyROA} we present the $R_{\rm BLR}$--$L_{5100}$ relation derived from the total sample as well as the best-quality subsample.  Based on $\tau_{\rm ICCF}$, the best-fit relation for the best-quality subsample is described as:
 
 \begin{align}
        &{\rm log} \, R_{\rm BLR} = 1.34 + 0.42\,({\rm log} \, L_{\rm 5100} - 43.89)  \label{eqa:R-L5100} 
 \end{align}
 
A consistent relation is found when we adopt $\tau_{\rm PyROA}$ in the analysis (see Table \ref{tab:ICCF_PyROA}). When we only include the best-quality AGNs, the slope the is still in agreement within $1\sigma$ uncertainty with that for the total  sample.   Our result is aligned with the findings by \citet{Woo24} who performed a similar analysis by collecting \hbeta\ lags from the literature \citep[see also][]{Malik23,Gravity24}. Our analysis confirms that the flattening of $R_{\rm BLR}$--$L_{5100}$ relation is real and remains consistent even after remeasuring \hbeta\ lags based on our uniform analysis and after selecting the best-quality subsample. 

\begin{table}[htbp]
    \centering
    \caption{$R_{\rm BLR}$--$L_{5100}$ Relations Using the Total and Best-Quality Subsample Based on ICCF and {\tt PyROA}}
    \setlength{\tabcolsep}{5pt}
    \begin{tabular}{ c c c c c c  }
    \hline \hline
    method  & sample   & $\beta$ & $\alpha$ &  $\sigma_{\rm int}$ \\ \hline
    
    \multirow{2}{*}{ICCF}  &  total   & $0.41^{+0.02}_{-0.02}$ & $1.33^{+0.02}_{-0.02}$ & $0.23^{+0.01}_{-0.01}$ \\ 
    &   best-quality  & $0.42^{+0.02}_{-0.02}$ & $1.34^{+0.02}_{-0.02}$ & $0.20^{+0.02}_{-0.02}$ \\ \hline

   \multirow{2}{*}{{\tt PyROA}}  & total  & $0.40^{+0.02}_{-0.02}$ & $1.29^{+0.02}_{-0.02}$ &  $0.25^{+0.01}_{-0.01}$ \\ 
            & best-quality  & $0.40^{+0.02}_{-0.02}$ & $1.30^{+0.02}_{-0.02}$ &  $0.24^{+0.02}_{-0.02}$ \\ 
    \hline
    \end{tabular}
    
    \label{tab:ICCF_PyROA}
\end{table}

We observe a slight decrease in $\sigma_{\rm int}$ when we only include AGNs with the best-quality  $\tau_{\rm ICCF}$,  e.g., from 0.23 to 0.20 dex. The  $\sigma_{\rm int}$ based on {\tt PyROA} doesn't exhibit significant changes which is understandable since our quality selection focuses on $\tau_{\rm ICCF}$.  In addition, the $\sigma_{\rm int}$ based on $\tau_{\rm PyROA}$ is $\sim0.03$ dex larger than that based on $\tau_{\rm ICCF}$, because of the smaller uncertainty of $\tau_{\rm PyROA}$. This demonstrates the influence of lag uncertainty on the $\sigma_{\rm int}$ of $R_{\rm BLR}$--$L_{5100}$ relation.  To our surprise, $\sigma_{\rm int}$ is only 0.25 dex even if we employ $\tau_{\rm PyROA}$ and use the total sample. While the intrinsic scatter is not as small as it was measured a decade ago by \citet{Bentz13}, the current $R_{\rm BLR}$--$L_{5100}$ relation is still  relatively tight, especially considering that the current RM sample is much larger and covers a broader range of Eddington ratio and luminosity. 

\begin{figure*}[htbp]
    \centering
   \includegraphics[width=0.85\textwidth]{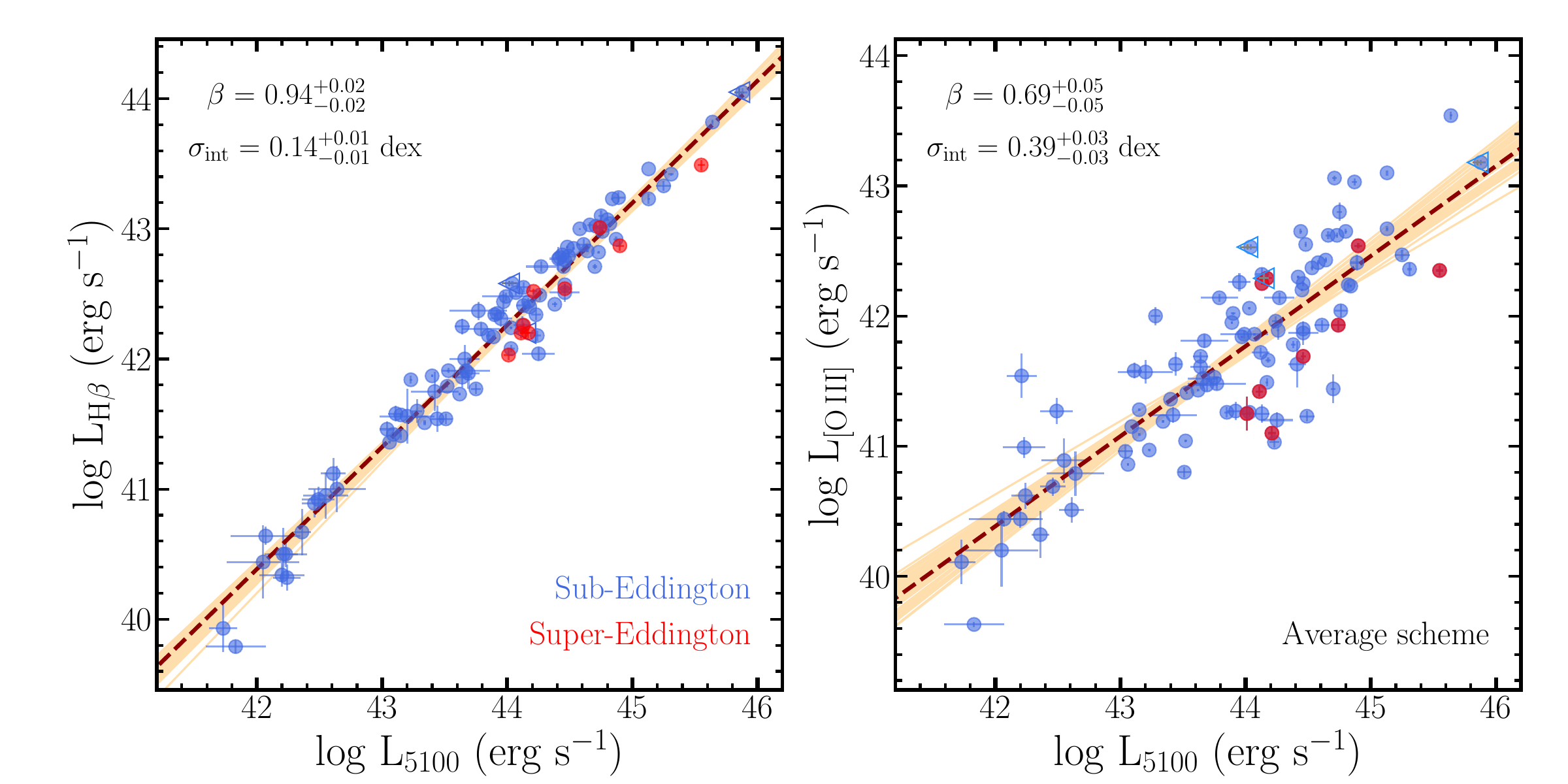}
    \caption{Comparison between different luminosity tracers in the average scheme. The left and right panel show the $L_{\rm H\beta}$--$L_{\rm 5100}$ and $L_{\rm [O\,\RNum{3}]}$--$L_{\rm 5100}$ relation, respectively. 
    Sub- and super-Eddington ratios are displayed using blue and red color, respectively.  
    The core-dominated radio-loud AGNs are highlighted by additional left-triangle envelopes. 
     The brown dashed lines represent the best-fit relations, while the orange lines are randomly selected 50 realizations from the MCMC samples for illustrating the dispersion of the relation.
     In each panel, the best-fit slope $\beta$ and the intrinsic scatter $\sigma_{\rm int}$ are labeled at the top left corner. } \label{fig:L-L_relation}
\end{figure*}

\section{The dependence on various luminosity tracers} \label{sec:luminositytracers}

In this section, we investigate the correlation of $R_{\rm BLR}$ with several luminosity tracers, i.e.,  the optical continuum luminosity $L_{5100}$, \hbeta\ luminosity $L_{\rm H\beta}$, and \OIII\ luminosity $L_{\rm [O\,\RNum{3}]}$. We use the best-quality subsample with $\tau_{\rm ICCF}$. 
In the case of AGNs with multiple lag measurements, we employ an average of them weighted by the uncertainties.
From the 157 best-quality lags, we obtain 101 distinct AGNs in the average scheme, for which all three luminosity measures are available.

\subsection{Luminosity Comparison} \label{sec:L-L_relation}

\begin{table}[htbp]
    \centering
    \caption{$L_{\rm H\beta}$--$L_{5100}$, and $L_{\rm [O\,III]}$--$L_{5100}$ Relations in Average Scheme}   \label{tab:L-L_relation}
        \setlength{\tabcolsep}{3pt}
    \begin{tabular}{ c c c c c }
    \hline \hline
    relation   & sample   & $\beta$ & $\alpha$ &   $\sigma_{\rm int}$ \\ \hline
         
    \multirow{2}{*}{$L_{\rm H\beta}$--$L_{5100}$}   &  total  & $0.94^{+0.02}_{-0.02}$ & $42.32^{+0.02}_{-0.02}$ & $0.14^{+0.01}_{-0.01}$ \\ 
    &   $\lambda_{\rm Edd}<1.0$ & $0.95^{+0.02}_{-0.02}$ & $42.33^{+0.02}_{-0.02}$ & $0.13^{+0.01}_{-0.01}$ \\ \hline

        \multirow{2}{*}{$L_{\rm [O\,\sevenrm{III}]}$--$L_{5100}$ }   & total  & $0.69^{+0.05}_{-0.05}$ & $41.82^{+0.04}_{-0.04}$ &  $0.39^{+0.03}_{-0.03}$ \\ 
          &   $\lambda_{\rm Edd}<1.0$ & $0.72^{+0.05}_{-0.05}$ & $41.84^{+0.04}_{-0.04}$ &  $0.38^{+0.03}_{-0.03}$  \\
    \hline
    \end{tabular}  
    \end{table}

First, we examine the relation between $L_{5100}$,  $L_{\rm H\beta}$, and $L_{\rm [O\,\RNum{3}]}$ in Figure \ref{fig:L-L_relation} by employing the same MCMC linear regression method described in \S \ref{sec:R-L_relation_ICCF_PyROA}. In this context, log$\,L_{5100}$ serves as an independent variable, while log$\,L_{\rm H\beta}$ and log$\,L_{\rm [O\,\RNum{3}]}$ act as a dependent variable. The pivot point, $x_0$, is set as the median value of log$\,L_{5100}$ for the total sample, i.e., 44.06 erg s$^{-1}$. We find the best-fit relation based on the total sample as:

\begin{align}
        &{\rm log} \, L_{\rm H\beta} = 42.32 + 0.94\,({\rm log} \, L_{\rm 5100} - 44.06)  \label{eqa:LHb-L5100}\\
       & {\rm log} \, L_{\rm [O\,\RNum{3}]} = 41.82 + 0.69\, ({\rm log} \, L_{\rm 5100} - 44.06) \label{eqa:LOIII-L5100}
\end{align}
The $L_{\rm H\beta}$--$L_{5100}$ relationship is characterized by a small intrinsic scatter, $\sigma_{\rm int}=0.14^{+0.01}_{-0.01}$ dex, indicating that host-corrected $L_{5100}$ is a good proxy for the Hydrogen ionizing continuum. 
The best-fit slope of $L_{\rm H\beta}$--$L_{5100}$ relation, $\beta=0.94^{+0.02}_{-0.02}$,  is statistically smaller (3$\sigma$) than unity, indicating the presence of a weak Baldwin effect of H$\beta$.  In contrast, the $L_{\rm [O\,\RNum{3}]}$--$L_{5100}$ relation exhibits a substantially larger intrinsic scatter and a notably shallower slope, compared to the $L_{\rm H\beta}$--$L_{5100}$ relation. 
It indicates a more pronounced Baldwin effect of \OIII\  relative to \hbeta, which is consistent with the results from the large sample of SDSS \citep{Shen14}.

We next investigate the influence of Eddington ratio by dividing our sample into sub- and super-Eddington AGNs. Super-Eddington AGNs exhibit systematically lower $L_{\rm H\beta}$ and $L_{\rm [O\, \RNum{3}]}$ compared to sub-Eddington AGNs in Figure \ref{fig:L-L_relation}. To quantify this deviation, we obtain the best-fit relation for sub-Eddington and super-AGN subsamples as summarized in Table \ref{tab:L-L_relation}. The median deviation of super-Eddington AGNs is $-0.21$ dex in $R_{\rm BLR}$ from the best-fit $L_{\rm H\beta}$--$L_{5100}$ relation of sub-Eddington AGNs, while that deviation is $-1.00$ dex for the $L_{\rm [O\, \RNum{3}]}$--$L_{5100}$ relation. Note that super-Eddington AGNs are mostly high-luminosity AGNs (Figure \ref{fig:L-L_relation}), therefore the derived slope of $L_{\rm H\beta}$--$L_{5100}$ relation  is slightly shallower than the value measured by \citet{DallaBonta20} (0.960$\pm$0.020), whose sample mostly consists of sub-Eddington AGNs. 

In addition, we explore other potential causes of the offset, for instance, the contamination of core-dominant radio-loud AGNs. Such objects may exhibit a large $L_{5100}$/$L_{\rm H\beta}$ ratio due to the substantial jet-induced contamination in the continuum \citep[e,g,][]{Greene05}. We compiled core-dominant radio-loud AGNs from literatures \citep[e.g.,][]{Brotherton96}. However, we find that these objects closely follow the $L_{\rm H\beta}$--$L_{5100}$ relation of the rest of the objects, suggesting the $R_{\rm BLR}$--$L_{5100}$ relation is not strongly biased due to such objects. 

\subsection{$R_{\rm BLR}$-- L Relation} \label{sec:R-L_relation}

\begin{figure*}[htbp]
    \centering
    \includegraphics[width=1.0\textwidth]{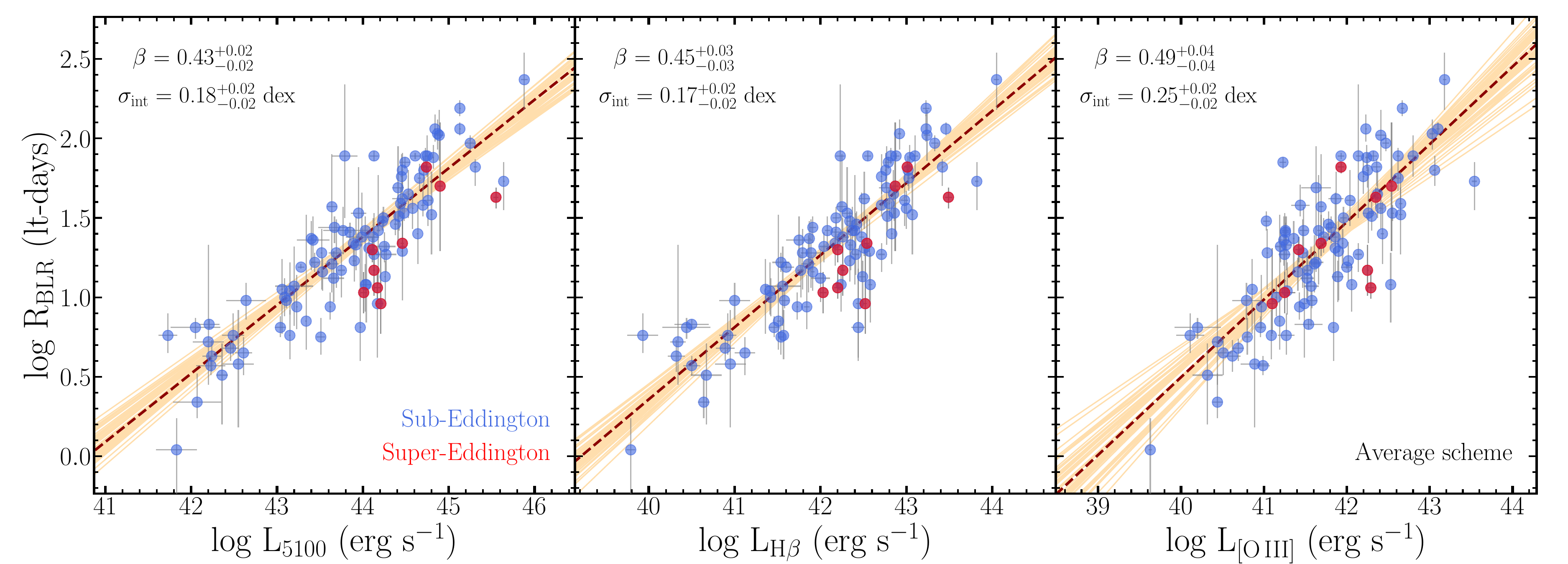}
    \caption{$R_{\rm BLR}$ as a function of different luminosity tracers in the average scheme, including $L_{5100}$ (left), $L_{\rm H\beta}$ (middle), and $L_{\rm [O\,\RNum{3}]}$ (right).  
    Sub- and super-Eddington objects are displayed using blue and red color, respectively. The brown dashed lines represent the best-fit relations of the total sample, and the orange lines are randomly selected 50 realizations from the MCMC samples for illustrating the dispersion of the relation. In each panel, the best-fit slope $\beta$ and the intrinsic scatter $\sigma_{\rm int}$ of the relation is presented at the upper-left corner.}
    \label{fig:R-L relation}
\end{figure*}

\begin{table*}
    \centering
    \caption{$R_{\rm H\beta}$--$L_{5100}$, $R_{\rm H\beta}$--$L_{\rm H\beta}$, and $R_{\rm H\beta}$--$L_{\rm [O\,III]}$ Relations in Average Scheme}
     \setlength{\tabcolsep}{10pt}
    \begin{tabular}{ c c c c c c  }
    \hline \hline
    Relation & Sample & $x_0$ & $\beta$ & $\alpha$ &  $\sigma_{\rm int}$ \\ \hline

     \multirow{2}{*}{$R_{\rm BLR}$--$L_{5100}$}   &  total  & 44.06 & $0.43^{+0.02}_{-0.02}$ & $1.41^{+0.02}_{-0.02}$ & $0.18^{+0.02}_{-0.02}$ \\ 

&  $\lambda_{\rm Edd}<1.0$  & 44.06 &  $0.45^{+0.02}_{-0.02}$ & $1.43^{+0.02}_{-0.02}$ & $0.16^{+0.02}_{-0.02}$ \\

         \multirow{2}{*}{$R_{\rm BLR}$--$L_{\rm H\beta}$}   &  total & 42.33 & $0.45^{+0.03}_{-0.03}$ & $1.42^{+0.02}_{-0.02}$ & $0.17^{+0.02}_{-0.02}$ \\ 
         
    &   $\lambda_{\rm Edd}<1.0$ & 42.33 & $0.46^{+0.03}_{-0.03}$ & $1.43^{+0.02}_{-0.02}$ & $0.16^{+0.02}_{-0.02}$ \\

    \multirow{2}{*}{$R_{\rm BLR}$--$L_{\rm [O\,\RNum{3}]}$}    & total & 41.69 &
    $0.49^{+0.04}_{-0.04}$ & $1.32^{+0.03}_{-0.03}$ & $0.25^{+0.02}_{-0.02}$ \\
    
    &   $\lambda_{\rm Edd}<1.0$ & 41.69 & $0.50^{+0.04}_{-0.04}$ & $1.33^{+0.03}_{-0.03}$ & $0.24^{+0.02}_{-0.02}$ \\  \hline
    

    \end{tabular}
    
    \label{tab:R-L_relation}
\end{table*}

Next, we investigate the $R_{\rm BLR}$ -- $L$ relation using three different luminosity tracers, i.e., $L_{5100}$, $L_{\rm H\beta}$, and   $L_{\rm [O\,\RNum{3}]}$.
 We obtain the best-fit relations in the average scheme as
 
\begin{align}
        &{\rm log} \, R_{\rm BLR} = 1.41 + 0.43\,({\rm log} \, L_{\rm 5100} - 44.06)  \label{eqa:R-L5100} \\
       & {\rm log} \, R_{\rm BLR} = 1.42 + 0.45\, ({\rm log} \, L_{\rm H\beta} - 42.33) \label{eqa:R-LHbeta}\\
       & {\rm log} \, R_{\rm BLR} = 1.32 + 0.49\, ({\rm log} \, L_{\rm [O\,\RNum{3}]} - 41.69) \label{eqa:R-LOIII}
\end{align}
with an intrinsic scatter, $0.18^{+0.02}_{-0.02}$, $0.17^{+0.02}_{-0.02}$, and $0.25^{+0.02}_{-0.02}$ dex, respectively for $R_{\rm BLR}$--$L_{5100}$,   $R_{\rm BLR}$--$L_{\rm H\beta}$,  and $R_{\rm BLR}$--$L_{\rm [O\,\RNum{3}]}$ relation (see Figure \ref{fig:R-L relation} and Table \ref{tab:R-L_relation}). 
The $R_{\rm BLR}$--$L_{\rm H\beta}$ relation shows the smallest intrinsic scatter as expected because the \hbeta\ line emission is a direct product of the BLR Hydrogen ionization and is less subject to defects in host corrections. It suggests that $L_{\rm H\beta}$ can serve as a good alternative to $L_{5100}$ for the future $R_{\rm BLR}$--$L$ relation analysis. On the other hand,  the $R_{\rm BLR}$--$L_{\rm [O\,\RNum{3}]}$ relation exhibits a larger intrinsic scatter, which is also expected because \OIII\ comes from the NLR, which covers a much larger scale. Consequently,  its variability time scale should also be much longer than that of broad emission lines \citep[e.g.,][]{Peterson13}.

As for the slope, we find that the slope varies when using different luminosity tracers. The $R_{\rm BLR}$--$L_{5100}$ relation displays a slope of $0.43^{+0.02}_{-0.02}$, while the $R_{\rm BLR}$--$L_{\rm H\beta}$ relation shows a slope of $\beta= 0.45^{+0.03}_{-0.03}$. The small difference is partly due to the weak \hbeta\ Baldwin effect, i.e., higher-luminosity AGNs have slightly lower $L_{\rm H\beta}$ / $L_{5100}$ ratios. In the case of the $R_{\rm BLR}$--$L_{\rm [O\,\RNum{3}]}$ relation, the best-fit slope is $\beta=0.49^{+0.04}_{-0.04}$. The derived slope is somewhat smaller than the previously reported slope based on a smaller \citep[$0.61\pm0.07$;][]{Greene10}.  
The difference is mainly related to the large difference in sample size. For instance, the inclusion of strong radio loud AGNs in our sample, e.g., PG 1100+772,  which may have enhanced jet-induced \OIII\ emission \citep{Mclntosh99}, results in a smaller slope.

Finally, we investigate the influence of Eddington ratio by comparing sub-Eddington and super-Eddington AGNs. Super-Eddington objects show substantial deviations in both $R_{\rm BLR}$--$L_{5100}$ and $R_{\rm BLR}$--$L_{\rm H\beta}$ relations (see Figure \ref{fig:R-L relation}).  These results are consistent with \citet{Du15}, who used a smaller sample \citep[see also][]{Woo24}. We quantify the deviation by obtaining the best-fit relation using only sub-Eddington AGNs (see Table \ref{tab:R-L_relation}).  The median deviation (in $R_{\rm BLR}$) of super-Eddington AGNs is $-0.29$ dex from the best-fit $R_{\rm BLR}$--$L_{5100}$ relation of sub-Eddington AGNs, while the deviation is $-0.21$ dex in the $R_{\rm BLR}$--$L_{\rm H\beta}$ relation. In contrast, super-Eddington AGNs offset by only $-0.07$ dex in the $R_{\rm [O\,\RNum{3}]}$--$L_{5100}$ relation. Consequently,  while the intrinsic scatter of the $L_{5100}$--$L_{\rm [O\,III]}$ relation is $\sim0.4$ dex, the $\sigma_{\rm int}$ of $R_{\rm BLR}$--$L$ relation increases by 0.07 dex when $L_{5100}$ is replaced with $L_{\rm [O\,III]}$.

\section{Discussion}
\subsection{Super-Eddington vs. Sub-Eddington AGNs} \label{sec:discussion1}

\begin{figure*}[htbp]
    \centering
     \includegraphics[width=0.9\textwidth]{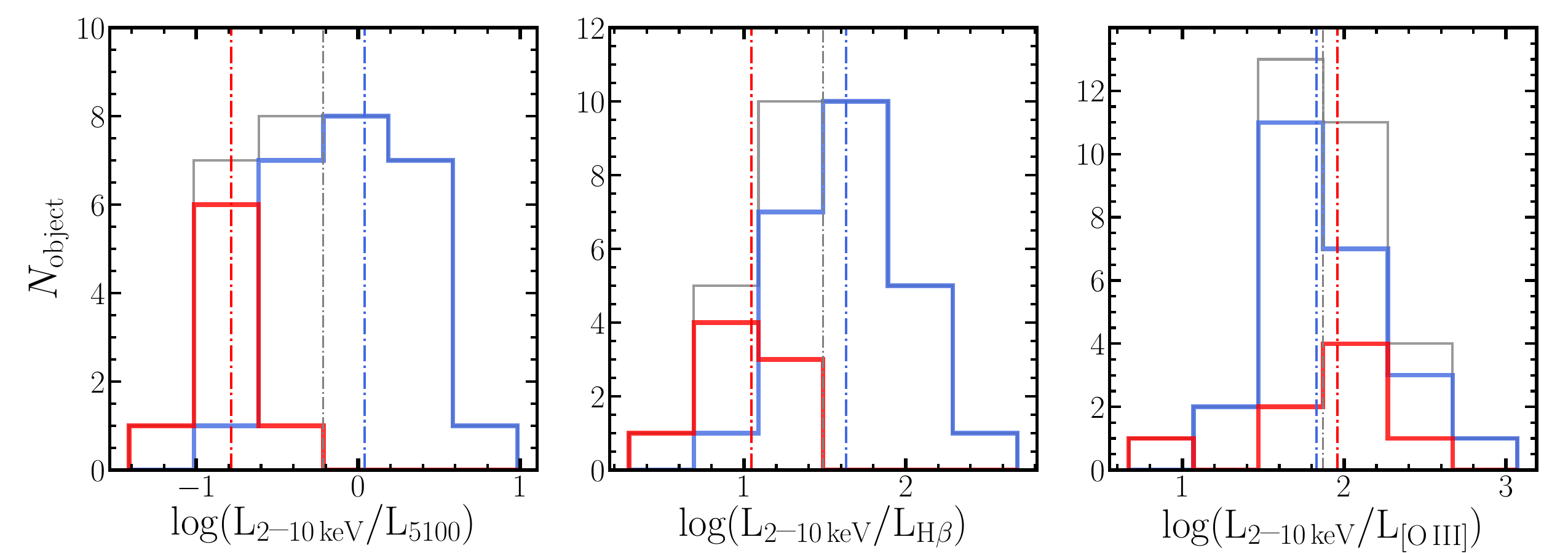}
    \caption{Histograms of the ratios  of hard X-ray luminosity  $L_{\rm 2\text{--}10\, keV}$ to $L_{5100}$ (left), $L_{\rm H\beta}$ (middle), and $L_{\rm [O\,III]}$ (right) based on a sample of 32 AGNs with hard X-ray luminosity available in \citet{Liu21}. 
    The blue and red histograms represent the distribution of low and high  $\lambda_{\rm Edd}$ subsample, respectively, with the division set as log$\lambda_{\rm Edd}\sim-0.3$. The vertical dashed lines in blue and red color indicate the median ratios for sub- and super-Eddington subsample, respectively. The grey histograms and vertical lines are the results for the total sample.}
    \label{fig:R-Xray0}
\end{figure*}

We confirm that super-Eddington AGNs show deviation in both $R_{\rm BLR}$--$L_{5100}$ and $R_{\rm BLR}$--$L_{\rm H\beta}$ relation (\S~\ref{sec:R-L_relation}), which is consistent with  \citet{Du15}. This deviation may be caused by various factors \citep[see the discussion by][]{Woo24}, including the self-shadowing effects associated with the slim disk \citep{Wang14}, the non-linear relation between optical and ionizing continua \citep{Fonseca-Alvarez20}, or the influence of BH spin \citep[e.g.,][]{Czerny19}.  We also demonstrate that  the $R_{\rm BLR}$--$L_{\rm [O\,III]}$ relation, on the contrary,  don't show such a deviation. In other words, for a given \OIII\ luminosity, sub-Eddington and super-Eddington AGNs do not  exhibit distinct $R_{\rm BLR}$ statistically. We discuss several possibilities that can lead to the different trends.

First, it could be due to the long timescale required for \OIII\ to respond to central variability, which could be $>>$10$^{3}$ years. Its luminosity only reflects an average accretion rate over a long period, during which AGNs are likely in a sub-Eddington phase for most of the time. Consequently, the \OIII\ luminosity of super-Eddington AGNs may under-represent the current accretion rate. This effect may compensate intrinsically shortened $R_{\rm BLR}$ in super-Eddington AGNs due to aforementioned factors, resulting in the lack of clear deviation between super- and sub-Eddington AGNs.

Second,  it could indicate that $R_{\rm BLR}$ is determined by higher-energy photons, given that the ionizing potential of \OIII\ is approximately 55 eV, while that for hydrogen is 13.6 eV.  This scenario, combined with the assumption that the SED in far UV (FUV) to extreme UV (EUV) range is softer in super-Eddington AGNs than in sub-Eddington AGNs, can naturally explain all observations, including the lack of deviation in $R_{\rm BLR}$--$L_{\rm [O\,\RNum{3}]}$ relation, and the presence of deviation in $R_{\rm BLR}$--$L_{\rm H\beta}$ relation,  and the even larger deviation in $R_{\rm BLR}$--$L_{5100}$ relation. However, examining the EUV SED is challenging due to observational limitations. If we seek theoretical models for explanation, we find that the slim disk model indeed predicts a saturation of emission at EUV wavelengths as the accretion rate increases above several times of the Eddington limits \citep[see Figure 8 in][]{Kubota19}. In contrast,  the FUV and optical emissions are less affected, resulting in a softer FUV to EUV SED in super-Eddington AGNs.

 \begin{figure*}[htbp]
    \centering
     \includegraphics[width=0.9\textwidth]{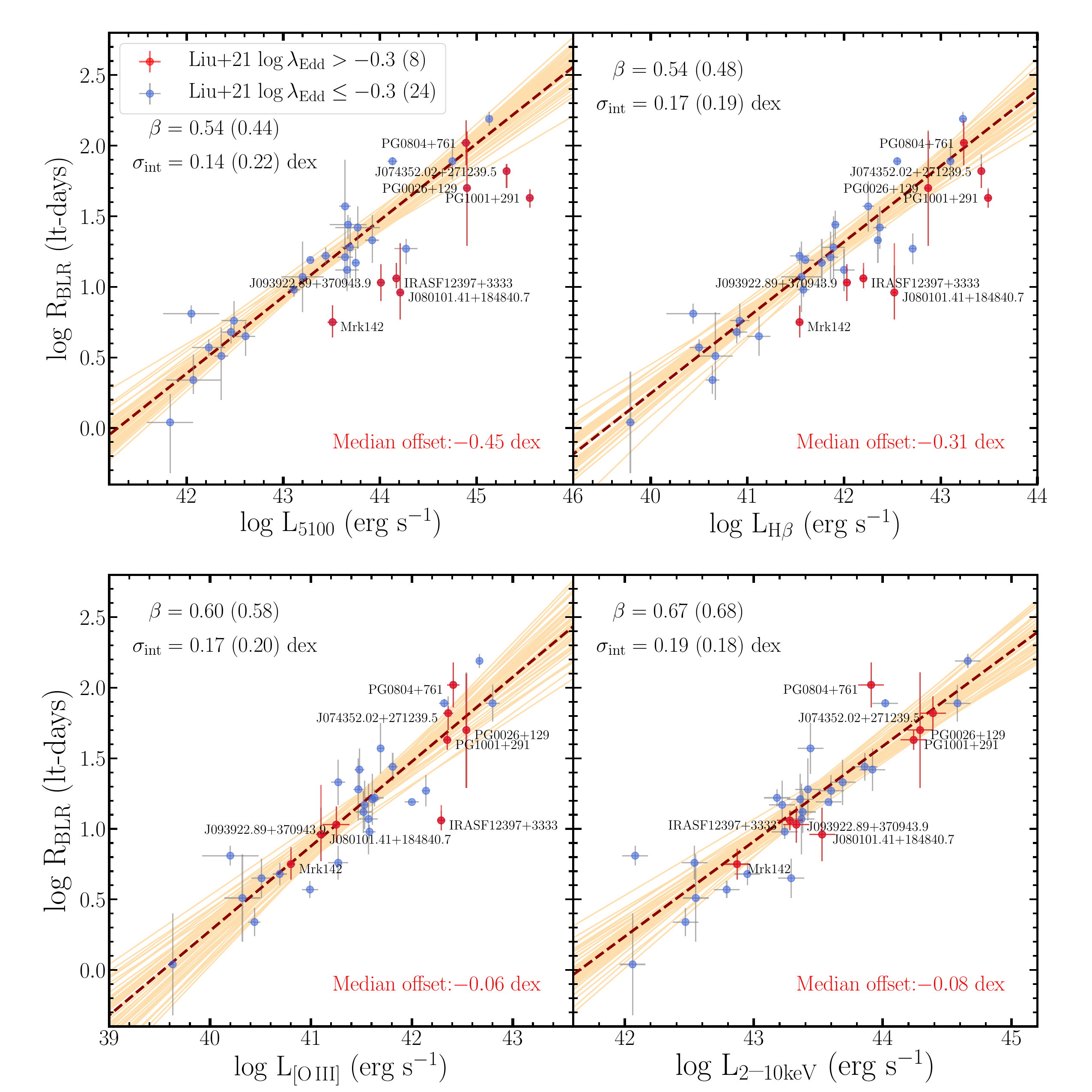}
    \caption{The $R_{\rm BLR}$--$L$ relation for the sample with X-ray luminosity measured by \citet{Liu21}.  The relation based on $L_{5100}$ (upper left), $L_{\rm H\beta}$ (upper right), $L_{\rm [O\,\RNum{3}]}$ (lower left), and $L_{\rm 2\text{--}10\, keV}$ (lower right) are presented. Twenty-four low and eight high-$\lambda_{\rm Edd}$ AGNs (divided by log$\lambda_{\rm Edd}=-0.3$) are shown in blue and red color, respectively. The name of each high $\lambda_{\rm Edd}$ AGN is given for comparing their different offset in different panels.  The brown dashed lines represent the best-fit relations of the sub-Eddington subsample while the orange lines are randomly selected 50 realizations from the MCMC samples for illustrating the dispersion of the relation. The best-fit slope $\beta$ and the intrinsic scatter $\sigma_{\rm int}$ for the low-$\lambda_{\rm Edd}$ subsample are presented at the upper-left corner, while the values in the brackets represent the results for the total sample. }
    \label{fig:R-Xray}
\end{figure*}

To distinguish these possibilities, we compile the 2 to 10 keV hard X-ray luminosity ($L_{\rm 2\text{--}10\,keV}$)  and study the $R_{\rm H\beta}$--$L_{\rm 2\text{--}10\,keV}$ relation. This idea is based on the observation that the hard X-ray emission in super-Eddington AGNs is also relatively weak \citep[e.g.,][]{Liu21}, while it is not subject to the long-timescale issue as for $L_{\rm [O\,\RNum{3}]}$.
In Figure \ref{fig:R-Xray0}, we present the distributions of the ratio between  $L_{\rm 2\text{--}10\,keV}$  and optical luminosities based on a sample of 32 AGNs, for which $L_{\rm 2\text{--}10\,keV}$ are available in \citet{Liu21}. Because of the limited sample size, we use log$\lambda_{\rm Edd}=-0.3$ to divide low-$\lambda_{\rm Edd}$  and high-$\lambda_{\rm Edd}$ subsamples. 
We find that high-$\lambda_{\rm Edd}$ AGNs exhibit smaller $L_{\rm 2\text{--}10\,keV}/L_{5100}$ and $L_{\rm 2\text{--}10\,keV}/L_{\rm H\beta}$ ratios compared to low-$\lambda_{\rm Edd}$ AGNs. The median differences are 0.83 dex and 0.59 dex for these two ratios, respectively. On the contrary, low-$\lambda_{\rm Edd}$ and high-$\lambda_{\rm Edd}$ AGNs only show a small difference, with a median of 0.13 dex, in the $L_{\rm 2\text{--}10\,keV}/L_{\rm [O\,\RNum{3}]}$ ratio.

 Figure \ref{fig:R-Xray}  presents the relation between $R_{\rm BLR}$ and $L_{\rm 2\text{--}10\,keV}$ for these 32 AGNs, compared to the relations based on other luminosity tracers.
Low and high-$\lambda_{\rm Edd}$ AGNs show only $0.08$ dex deviation in the $R_{\rm BLR}$--$L_{\rm 2\text{--}10\,keV}$ relation, which is similar to the case of $L_{\rm [O\,III]}$. 
We also find that the $\sigma_{\rm int}$ of $R_{\rm BLR}$--$L_{\rm 2\text{--}10\,keV}$ relation remains consistent with or without including high-$\lambda_{\rm Edd}$ AGNs (see Figure \ref{fig:R-Xray}). 
On the contrary, the deviation is substantial in the $R_{\rm BLR}$--$L_{5100}$ relation  as well as the $R_{\rm BLR}$--$L_{\rm H\beta}$  relation (similar to the results presented in \S \ref{sec:R-L_relation} based on larger sample). Regarding the slopes, all of the slopes are larger than those reported in Figure \ref{fig:R-L relation} based on larger sample, which is mainly due to the large difference in sample size and potential sample selection effect.

These results can be considered as evidence of the EUV SED scenario, given that the hard X-ray emission in AGNs is mainly originated from EUV photons  through inverse Comptonization.  In other words, the same EUV photons that are responsible for generating hard X-ray emission could also determine $R_{\rm BLR}$. The larger-than-0.5 slope, if confirmed, could be resulted from the non-linear interaction between the disk and corona \citep[e.g.,][]{Lusso10}, where more luminosity AGNs tend to have smaller hard X-ray to optical/UV flux ratios.  The results from Figure \ref{fig:R-Xray} actually reveal a connection among the hard X-ray emission from the innermost region, the BLR sizes, and \OIII\ emission from the narrow line region, which cannot be easily explained by the properties of a single component,  such as BLR density or  BLR emission anisotropy. The ionizing SED provides a  natural explanation.

The recent results from GRAVITY show that both the torus size -- $L$ relation and BLR size -- $L$ relation show a slope of $\sim$0.5 \citep{Gravity24,Gravity24b} using hard X-ray luminosity,  while they exhibit a slope of $\sim$ 0.4 using $L_{5100}$ \citep[e.g.,][]{Gravity23,Gravity24b, Chen23,Mandal24}. While the slope is smaller than what we measured here due to either sample selection effects or systematics between RM and interferometric observations, the results demonstrate the critical role of the ionizing SED in determining the slope of the $R_{\rm BLR}$–$L$ relation.

While the $L_{\rm 2\text{--}10\, keV}$ is not subject to timescale issues,  it is not measured simultaneously with $R_{\rm BLR}$. The large variability of X-ray can result in a large scatter and erase any small deviation.  Nonetheless, we find that the $\sigma_{\rm int}$ of the $R_{\rm BLR}$--$L_{\rm 2\text{--}10\,keV}$ relation is not large (0.18 dex,  if assuming the uncertainty of $L_{\rm 2\text{--}10\,keV}$ as 0.10 dex).  \citet{Liu21} only adopted high state X-ray observations, which could be a potential reason for the small scatter. It is also possible that with larger sample size, the $\sigma_{\rm int}$ will increase.  The $R_{\rm BLR}$--$L_{\rm 2\text{--}10\,keV}$ relation should be tested in the future with quasi-simultaneous X-ray and UV-optical observations.

\subsection{Future Improvement}

In this work, we investigated the influence of lag measurements and  the choice of luminosity tracer on the $R_{\rm BLR}$--$L$ relation. We discuss several future directions, which could improve the $R_{\rm BLR}$--$L$ relation.

First, since the correction for the host contamination has not been homogeneously performed in measuring $L_{5100}$,
a future direction is to perform uniform host decomposition for the entire RM sample using high resolution images, e.g., from the Hubble space telescope.  This is difficult given the rapidly growing RM sample size. 
The use of $L_{\rm H\beta}$ instead of $L_{5100}$ may be a good alternative for studying the $R_{\rm BLR}$ -- $L$ relation. However, the bias due to the Eddington ratio still affects the $R_{\rm BLR}$--$L_{\rm H\beta}$  relation.

Second, while the absence of the rest-frame extreme UV observations hinders understanding of AGN ionizing SED
multi-band simultaneous observations may help improving our understanding of AGN SED. By combining quasi-simultaneous optical, near-UV, and soft X-ray observation, one may obtain relatively good constraints on the ionizing SED shape. Expanding these constraints to a larger RM sample, 
the systematic trend in the $R_{\rm BLR}$--$L$ relation can be better understood. 
Alternatively, studying the far/near-UV to optical photometry of a large sample of high-redshift ($z>2$) AGNs can also provide useful constraints \citep[e.g.,][]{Zheng97, Krawczyk13}. 
    
Third, new RM campaigns can be carried out for AGNs with previous lag measurements, especially for those from early years with moderate or low cadence. The lags of some super-Eddington AGNs are suspected to be biased by their insufficient cadence (e.g., PG 0804+761). Thus, new campaign can provide consistency check and may reduce the scatter of the $R_{\rm BLR}$--$L$ relation.

\section{Summary} \label{sec:summary}

In this paper we investigated the H$\beta$ lag measurements and the BLR size-luminosity relation using a large sample of AGNs. We summarize the main results as follows:
\begin{enumerate}

 \item We remeasure the \hbeta\ lags for 212 AGNs in the literature using three different methods, i.e., ICCF, {\tt JAVELIN} and {\tt PyROA}. ICCF is the most reliable approach albeit with larger uncertainties compared to other methods. While $\tau_{\rm JAVELIN}$ is consistent with $\tau_{\rm ICCF}$ for most objects, it fails for some objects as the cross correlation result is trapped in the seasonal gaps or not converged.  
In the case of {\tt PyROA} we obtained generally consistent lags but with smaller lag uncertainties compared to those of ICCF. 
    
        \item We obtain the best-fit $R_{\rm BLR}$--$L_{5100}$ relation based on ICCF as ${\rm log} \, R_{\rm BLR} = 1.34 + 0.42\,({\rm log} \, L_{\rm 5100} - 43.89)$,  with an intrinsic scatter of 0.20$^{+0.02}_{-0.02}$ dex,  when combining the best-quality ICCF lags from literature sample and SAMP. Compared to the results without quality selection, the slope is consistent but the intrinsic scatter is slightly reduced.  When $\tau_{\rm PyROA}$ is used, we find generally consistent results with those based on ICCF but with slightly larger intrinsic scatter ($\sim $0.24 dex) due to its smaller lag uncertainty.

    \item We investigate different luminosity tracers in the $R_{\rm BLR}$--$L$ relation. We find that \hbeta\ show a weak Baldwin effect, while \OIII\ exhibit a significant Baldwin effect. Super-Eddington AGNs have lower $L_{\rm H\beta}$ and $L_{\rm [O\,III]}$ at given $L_{5100}$, while the deviation is stronger in the $L_{\rm [O\,III]}$--$L_{5100}$ relation.
    
    \item In the average scheme, the best-fit slope increases from the $0.43^{+0.02}_{-0.02}$ for the $R_{\rm BLR}$--$L_{5100}$ relation, to $0.45^{+0.03}_{-0.03}$ for the $R_{\rm BLR}$--$L_{\rm H\beta}$ relation, and to $0.49^{+0.04}_{-0.04}$ for $R_{\rm BLR}$--$L_{\rm [O\,\RNum{3}]}$ relation. The $R_{\rm BLR}$--$L_{\rm H\beta}$ relation exhibits a small $\sigma_{\rm int}$ ($\sim$0.17 dex), while the $R_{\rm BLR}$--$L_{\rm [O\,\RNum{3}]}$ relation shows a larger $\sigma_{\rm int}$ ($\sim$0.25 dex). The deviation of super-Eddington AGNs is significant in both $R_{\rm BLR}$--$L_{5100}$ and $R_{\rm BLR}$--$L_{\rm H\beta}$ relations, while it is insignificant in the $R_{\rm BLR}$--$L_{\rm [O\,\RNum{3}]}$ relation.

   \item We discuss several possibilities for the insignificant deviation between sub- and super-Eddington AGNs in the $R_{\rm BLR}$--$L_{\rm [O\,\RNum{3}]}$ relation,  including the \OIII\ variability time scale, the difference in the FUV to EUV SED. Based on a small sample of AGNs with $L_{\rm 2\text{--}10\,keV}$ available, we find that the $R_{\rm BLR}$--$L_{\rm 2\text{--}10\,keV}$ relation also exhibits no significant difference between low- and high-Eddington ratio subsample (see \S \ref{sec:discussion1} for details). Such observations could be considered as evidence supporting the ionizing SED scenario.

\end{enumerate}


This work has been supported by the Basic Science Research Program through the National Research Foundation of Korean Government (2019R1A6A1A10073437 and 2021R1A2C3008486). We thank the anonymous referee for the helpful comments and suggestions. We thank Hengxiao Guo for initial light curve collection. We thank Donghoon Son, Hojin Cho, Amit Kumar Mandal, Jaejin Shin, Hengxiao Guo for lag quality assessment. 

\clearpage 
\newpage

\appendix
\section{Lag measurements} \label{sec:appendixA}

\renewcommand\thefigure{\thesection.\arabic{figure}} 
\setcounter{figure}{0}
\renewcommand\thetable{\thesection.\arabic{table}} 
\setcounter{table}{0}

Table \ref{tab:Lag_measurement} records the lag measurements and related parameters for the literature parent sample, while \ref{tab:Lag_measurement_SAMP} summarizes the lag measurements for SAMP.  Table \ref{tab:average_lags} summarizes the lags and luminosities in the average scheme.

\begin{longtable*}[h]{ p{0.5cm}   p{4cm}  p{1.0cm}  p{1.0cm}  p{10cm}  }
\caption{Lag measurements and quality parameters for literature parent sample}\label{tab:Lag_measurement}\\
\hline \hline \\[-2.3ex]
   {\textbf{No.}} &
   {\textbf{Column}} &
   {\textbf{Format}} &
   {\textbf{Unit}} &
   {\textbf{Description}} \\ [0.5ex] \hline
   \\[-2.3ex]
\endfirsthead
1	&ID	     	        		&Long           &               &Object ID  \\	
2	&Object    	        		&String          &               &Object Name  \\	
3	&Reference      		&String          &               &Reference paper of the light curve and initial lag measurement  \\	
4      &Redshift                   &Long           &               &Redshift of the object \\ 
5      &Rmax                        &Long           &               &Maximum correlation coefficient $r_{\rm max}$ \\
6      &Rmax\_DE                 &Long           &               &Maximum correlation coefficient $r_{\rm max}$ after detrending \\
7      &p-value                    &Long           &               &$p$-value  from a cross-correlation reliability test\\
8	&TAU\_LITERATURE    		&Long           & Days             & Literature reported rest-frame lag measurement; $-1$ indicates not available \\	
9	&e\_TAU\_LITERATURE	&Long           & Days             & Lower uncertainty of  TAU\_LITERATURE   \\	
10	&E\_TAU\_LITERATURE	&Long           & Days             & Upper uncertainty of  TAU\_LITERATURE   \\	
11	&TAU\_ICCF    		&Long           & Days             & Final rest-frame lag measurement based on ICCF centroids  \\	
12	&e\_TAU\_ICCF	&Long           & Days             & Lower uncertainty of  TAU\_ICCF   \\	
13	&E\_TAU\_ICCF	&Long           & Days             & Upper uncertainty of  TAU\_ICCF   \\	
14    &Fpeak\_ICCF       	&Long           &                      & $f_{\rm peak}$ of  TAU\_ICCF \\
15      &Flag\_Detrend\_ICCF       	&Long           &                      & Flag for detrending: 1 means detrended while 0 means not detrended \\
16      &Flag\_Posterior\_ICCF       	&Long           &                      & Flag on posterior distributions for ICCF centroids. 1 means good; 2 means not-converged; 3 means the primary peak is affected by the boarder of the lag searching window. \\
17	&Visual\_Inspection\_ICCF	&Long   &                     & Visual quality assessment of the TAU\_ICCF, where 5 means the best quality, while 1 means less good quality.  \\
18	&TAU\_JAVELIN    	&Long           & Days             & Final rest-frame lag measurement based on  {\tt JAVELIN} \\	
19	&e\_TAU\_JAVELIN	&Long           & Days             & Lower uncertainty of  TAU\_JAVELIN   \\	
20	&E\_TAU\_JAVELIN	&Long           & Days             & Upper uncertainty of  TAU\_JAVELIN   \\
21      &Fpeak\_JAVELIN	&Long           &                      & $f_{\rm peak}$ of  TAU\_JAVELIN \\
22      &Flag\_Posterior\_JAVELIN       	&Long           &                      & Flag on posterior distributions for JAVELIN. Same as Flag\_Posterior\_ICCF.   \\
23	&TAU\_PyROA    	&Long           & Days             & Final rest-frame lag measurement based on {\tt PyROA}  \\	
24	&e\_TAU\_PyROA	&Long           & Days             & Lower uncertainty of  TAU\_PyROA   \\	
25	&E\_TAU\_PyROA	&Long           & Days             & Upper uncertainty of  TAU\_PyROA   \\
26      &Fpeak\_PyROA	&Long           &                      & $f_{\rm peak}$ of  TAU\_PyROA \\
27      &Flag\_Posterior\_PyROA       	&Long           &                      & Flag on posterior distributions for PyROA. Flag\_Posterior\_ICCF. \\
28	&Flag\_Best\_Quality	&Long           &                      & Equal 1: Selected; Equal 0: Not selected   \\	   	
29      &LogL5100	&Long           &  erg s$^{-1}$           & Log luminosity at 5100\AA\ \\
30      &e\_LogL5100	&Long           &   erg s$^{-1}$           & Uncertainty of  LogL5100\\
31      &FWHM\_mean 	&Long           &  km s$^{-1}$           & Collected FWHM from mean spectra; $-1$ indicates not available \\
32      &e\_FWHM\_mean	&Long           &   km s$^{-1}$           & Uncertainty of  FWHM\_mean\\
33      &Sigma\_rms	&Long           &  km s$^{-1}$           & Collected $\sigma_{\rm line}$ from rms spectra; $-1$ indicates not available \\
34      &e\_Sigma\_rms	&Long           &   km s$^{-1}$           & Uncertainty of  Sigma\_rms\\
\hline
\end{longtable*}

\begin{longtable*}[h]{ p{0.5cm}   p{4cm}  p{1.0cm}  p{1.0cm}  p{10cm}}
\caption{Lag measurements and quality parameters for SAMP sample}\label{tab:Lag_measurement_SAMP}\\
\hline \hline \\[-2.3ex]
   {\textbf{No.}} &
   {\textbf{Column}} &
   {\textbf{Format}} &
   {\textbf{Unit}} &
   {\textbf{Description}} \\ [0.5ex] \hline
   \\[-2.3ex]
\endfirsthead
1	&Object    	        		&String          &               &Object Name  \\	
2      &Redshift                   &Long           &               &Redshift of the object \\ 
3	&TAU\_ICCF    		&Long           & Days             & Final observed-frame lag measurement based on ICCF (centroids)  \\	
4	&e\_TAU\_ICCF	&Long           & Days             & Lower uncertainty of  TAU\_ICCF   \\	
5	&E\_TAU\_ICCF	&Long           & Days             & Upper uncertainty of  TAU\_ICCF   \\	
6      &Fpeak\_ICCF       	&Long           &                      & $f_{\rm peak}$ of  TAU\_ICCF \\
7	&TAU\_JAVELIN    	&Long           & Days             & Final observed-frame lag measurement based on  {\tt JAVELIN} \\	
8	&e\_TAU\_JAVELIN	&Long           & Days             & Lower uncertainty of  TAU\_JAVELIN   \\	
9	&E\_TAU\_JAVELIN	&Long           & Days             & Upper uncertainty of  TAU\_JAVELIN   \\
10      &Fpeak\_JAVELIN	&Long           &                      & $f_{\rm peak}$ of  TAU\_JAVELIN \\
11	&TAU\_PyROA    	&Long           & Days             & Final observed-frame lag measurement based on {\tt PyROA}  \\	
12	&e\_TAU\_PyROA	&Long           & Days             & Lower uncertainty of  TAU\_PyROA   \\	
13	&E\_TAU\_PyROA	&Long           & Days             & Upper uncertainty of  TAU\_PyROA   \\
14     &Fpeak\_PyROA	&Long           &                      & $f_{\rm peak}$ of  TAU\_PyROA \\
15	&Flag\_Best\_Quality\_ICCF	&Long           &               & Flag indicates the final selected best-quality sample based on ICCF. Equal 1: selected best-quality ones; Equal 0: not selected.   \\	   	
16      &LogL5100	&Long           &  erg s$^{-1}$           & Log luminosity at 5100\AA\ \\
17      &e\_LogL5100	&Long           &   erg s$^{-1}$           & Uncertainty of  LogL5100\\

\hline
\end{longtable*}

\begin{longtable*}[h]{ p{0.5cm}   p{4cm}  p{1.0cm}  p{1.0cm}  p{10cm}}
\caption{Lags and Luminosities in the Average Scheme} \label{tab:average_lags}\\
\hline \hline \\[-2.3ex]
   {\textbf{No.}} &
   {\textbf{Column}} &
   {\textbf{Format}} &
   {\textbf{Unit}} &
   {\textbf{Description}} \\ [0.5ex] \hline
   \\[-2.3ex]
\endfirsthead
1	&Object    	        		&String          &               &Object Name  \\	
2      &Redshift                   &Long           &               &Redshift of the object \\ 
3	&LogR    		&Long           & lt-days             & Log BLR Radius  \\	
4	&e\_LogR 	&Long           & lt-days             & Lower uncertainty of  LogR   \\	
5	&E\_LogR	        &Long           & lt-days             & Upper uncertainty of  LogR   \\	
6      &LogL5100	&Long           &  erg s$^{-1}$           & Log AGN luminosity at 5100\AA\ \\
7      &e\_LogL5100	&Long           &   erg s$^{-1}$           & Uncertainty of  LogL5100\\
8      &LogLHbeta	&Long           &  erg s$^{-1}$           & Log Luminosity of Broad \hbeta; $-1$ indicates not available \\
9      &e\_LogLHbeta 	&Long           &   erg s$^{-1}$           & Uncertainty of  LogLHbeta\\
10      &LogLOIII	&Long           &  erg s$^{-1}$           & Log luminosity of \OIII; $-1$ indicates not available\\
11     &e\_LogLOIII	&Long           &   erg s$^{-1}$           & Uncertainty of  LogLOIII\\
12      &LogL2-10keV	&Long           &  erg s$^{-1}$           & Log Hard X-ray luminosity at 2-10 keV; $-1$ indicates not available \\
13	&LogLambda\_Edd    		&Long           &              & Log Eddington Ratio  \\	
14	&e\_LogLambda\_Edd 	&Long           &              & Lower uncertainty of  LogLambda\_Edd    \\	
15	&E\_LogLambda\_Edd	        &Long           &             & Upper uncertainty of  LogLambda\_Edd    \\

\hline
\end{longtable*}

\bibliography{ref}
\end{document}